%% file: haggies-cpc.tex
\documentclass[preprint,number,12pt]{elsarticle}
\pdfoutput=1

\usepackage{amsmath}
\usepackage{dsfont}
\usepackage{listings}
\usepackage{graphicx}
\usepackage[thinspace,mediumqspace,squaren,textstyle]{SIunits}
\usepackage{color}
\usepackage[printonlyused]{acronym}
\usepackage[pdftex]{hyperref}
\usepackage{algorithmic}
\usepackage{algorithm}
\usepackage{longtable}

\addunit{\byte}{B}

\journal{Computer Physics Communications}

\newcommand{\xfigbox}[2]{\includegraphics[angle=0,width=#1]{#2}}

\input lstform.tex
\input lsthaggies.tex

\lstnewenvironment{form}{%
	\lstset{language=form,%
		frame=single,frameround=tttt,framexleftmargin=6pt,%
		indexstyle=[1]\indexprocedures,%
		indexstyle=[2]\indexsymbols,%
		indexstyle=[3]\indexfunctions%
	}%
}{}

\newcommand{\FORM}{{\texttt{Form}}}
\newcommand{\FortranXC}{{\texttt{Fortran~90}}}
\newcommand{\GOLEM}{{\texttt{Golem}}}
\newcommand{\haggies}{{\texttt{haggies}}}

\newcommand{\kea}[1]{\vert #1 \rangle}

\newcommand{\Spaa}[1]{\langle #1 \rangle}

\newcommand{\Spbb}[1]{[ #1 ]}

\hypersetup{
	pdftitle={Optimising Code Generation with `haggies'},
	pdfauthor={Thomas Reiter},
	pdfkeywords={Computer algebra, Code generation, %
		Automation of perturbative calculations},
	pdfsubject={Generation of optimised programs for the %
		evaluation of possibly large algebraic expressions}
}

\begin{document}
\begin{frontmatter}
\title{\vspace{-2.5cm}\hfill {\small\rm Nikhef-2009-012}\vspace{1cm}\\
Optimising Code Generation with \haggies}

\author{T.~Reiter}
\address{Nikhef, Science Park 105, 1098 XG Amsterdam,
             The Netherlands}
\ead{thomasr@nikhef.nl}
\ead[url]{http://www.nikhef.nl/\~{}thomasr/}

\begin{abstract}
This article describes \haggies, a program for the generation
of optimised programs for the efficient numerical evaluation
of mathematical expressions. It uses a multivariate Horner-scheme
and \acl{CSE} to reduce the overall number of operations.

The package can serve as a back-end for
virtually any general purpose computer algebra program.
Built-in type inference that
allows to deal with non-standard data types in strongly typed
languages and a very flexible, pattern-based output specification
ensure that \haggies{} can produce code for a large variety
of programming~languages.

We currently use \haggies{} as part of an automated package for
the calculation of one-loop scattering amplitudes in
quantum field theories. The examples in this articles, however,
demonstrate that its use is not restricted to the field of
high energy~physics.
\end{abstract}

\end{frontmatter}

{\bf PROGRAM SUMMARY}

\begin{small}
\noindent
{\em Manuscript Title:}
Efficient Programs from Expressions with \haggies\\
{\em Authors:} T.~Reiter\\
{\em Program Title:} \haggies\\
{\em Programming language:} Java, JavaCC \\
{\em Operating system:}
  Any system that runs the Java Virtual Machine\\
{\em RAM:} determined by the size of the problem \\
{\em Number of processors used:}
dynamical, up to the number of installed CPUs 
{\em Classification:}
	4.14 Utility,
   5 Computer Algebra,
	6.2 Languages,
	6.5 Software including Parallel Algorithms,
   11.1  General, High Energy Physics and Computing\\
{\em Nature of problem:} Generation of optimised programs for the
  evaluation of possibly large algebraic expressions\\
%
{\em Solution method:} Java implementation\\
\end{small}

\newpage

\hspace{1pc}
{\bf LONG WRITE-UP}

\section{Motivation}\label{sec:motivation}
In physics problems with large numbers of parameters the appearance of
large expressions is commonly observed.
Exploiting the symmetries of the problem often simplifies the expression
sufficiently to allow for a direct numerical implementation.
However, these simplifications sometimes are not obvious 
due to the lack of an algorithmic description in which order
to apply the underlying relations.
In this case they are not suitable for an automated~setup.

One representative class of problems are higher order corrections in
perturbative quantum field theory. In order to meet the precision of
modern collider experiments such as the \ac{LHC},
the leading order approximation of
scattering amplitudes is often not sufficient and at least one-loop precision
is required for many processes with up to four particles in the final
state.~\cite{Bern:2008ef}. It is therefore not surprising that recently
the automation of such calculations has become a very active field
of research, leading to new tools and promising techniques~\cite{%
Berger:2008ag,Binoth:2008uq,%
Bredenstein:2008ia,Diakonidis:2009fx,%
Fujimoto:2008zz,Giele:2008bc,Hahn:2000kx,Hahn:2006qw,%
Lazopoulos:2008ex,Reiter:2009kb,vanHameren:2009dr,vanHameren:2009vq,%
Winter:2009kd}. The interested reader will find
a more general overview over the current status of higher order corrections
in particle physics for example 
in~\cite{Binoth:2009fk} or in~\cite{Bern:2008ef}.

The calculation of scattering amplitudes to one-loop
precision can be divided into two subproblems. One part of the amplitude,
the real corrections, consists of tree-level Feynman diagrams,
describing the radiation of an extra, unobserved particle.
The second half of the calculation describes the exchange of
a virtual particle and leads to one-loop diagrams. Both parts of the
calculation contain singularities which cancel after summing over both
of the two contributions.
In \ac{QCD}, a commonly used technique for the regularisation
of these singularities is the subtraction method
by Catani and~Seymour~\cite{Catani:1996vz}, which has led to
several automated implementations~\cite{%
Gleisberg:2007md,Seymour:2008mu,%
Hasegawa:2008ae,Frederix:2008hu,Czakon:2009ss}.
Combined with existing tools for tree-level
calculations the real corrections, these implementations
provide a complete solution for the real emission corrections.

The calculation of the virtual corrections requires the computation
of one-loop Feynman diagrams. The two basic strategies are a fully
numerical approach, which is mainly followed in the implementation
of unitarity-based methods~\cite{%
Berger:2008ag,Giele:2008bc,vanHameren:2009dr},
and a semi-numerical/algebraic approach which appears to be well-suited
for calculations based on Feynman diagrams~\cite{Binoth:2008uq,%
Fujimoto:2008zz,Hahn:2000kx,Hahn:2006qw,%
Reiter:2009kb}.

\section{Background}
Here, the package \GOLEM{} is discussed,
which uses the latter technique to produce a numerically stable
representation of the virtual corrections of cross-sections.
This package forms the original motivation for the development of~\haggies{}.

The matrix element for the virtual corrections to a process in \ac{QCD}
with $N$ particles in the final state can be written as
\begin{equation}
\mathcal{M}_{\text{virt}}=\sum_{\kea{c}\in\mathcal{B}}
\mathcal{A}_c(p_a,p_b;p_1,\ldots,p_N)\cdot\kea{c}\text{,}
\end{equation}
where $\mathcal{B}$ denotes some basis for the colour tensor
of the external particles. The right hand side can be 
decomposed further by projecting onto helicity states and by using the
fact that the matrix element is defined as the sum over all contributing
Feynman diagrams $\mathcal{G}$,
\begin{align}
\mathcal{M}_{\text{virt}}^{\{\lambda\}}&=
\sum_{\mathcal{G}}
\sum_{\kea{c}\in\mathcal{B}}
\mathcal{G}_c(p_a^{\lambda_a},p_b^{\lambda_b};%
p_1^\lambda,\ldots,p_N^{\lambda_N})\cdot\kea{c}\text{,}\\
\left\vert\mathcal{M}\right\vert^2&=
\sum_{\{\lambda\}}
\mathcal{M}_{\text{virt}}^{\{\lambda\}}
\left(\mathcal{M}_{\text{virt}}^{\{\lambda\}}\right)^\dagger\text{.}
\end{align}

One of the tasks of the \GOLEM{} implementation is to generate a
\FortranXC{} subroutine for each diagram
$\mathcal{G}_c(p_a^{\lambda_a},p_b^{\lambda_b};%
p_1^\lambda,\ldots,p_N^{\lambda_N})\cdot\kea{c}$. Each of these
diagrams is a linear combination of tensor integrals of the type
\begin{equation}\label{eq:define-tensint}
I^{n;\mu_1\ldots\mu_r}_N(S)=\int\!\!\frac{\mathrm{d}^nk}{i\pi^{n/2}}
\frac{k^{\mu_1}\cdots k^{\mu_r}}{\prod_{j=1}^{N}\left[(k+r_j)^2-m_j^2\right]}
\end{equation}
contracted with corresponding coefficients $c_{\mu_1\ldots\mu_r}$.
The vectors $r_j$ are linear combinations of the external momenta and
$m_j$ are the propagator masses. The integrals are dimensionally regularised
in $n=(4-2\varepsilon)$ dimensions and the result can be expressed as a
Laurent series $A/\varepsilon^2+B/\varepsilon+C+\mathcal{O}(\varepsilon)$.
The algorithm described in~\cite{Binoth:2008uq} is used
to express the tensor integrals in terms of Feynman parameter integrals
of the form
\begin{multline}\label{eq:fpintegral}
I^{d}_N(l_1,\ldots,l_p;S)=(-1)^N\Gamma(N-d/2)\times\\
\int\prod_{j=1}^N\left(\mathrm{d}z_j\,\Theta(z_j)\right)
\delta\left(1-\sum_i z_i\right)\frac{z_{l_1}\cdots z_{l_p}}{%
\left(-\frac12z^{\mathsf T}Sz-i\delta\right)^{N-d/2}}
\end{multline}
with the matrix
\begin{equation}
S_{ij}=(r_i-r_j)^2-m_i^2-m_j^2\text{.}
\end{equation}
The coefficients in front of these integrals are in general
rational polynomials
in terms of spinor products of external vectors ($\Spaa{p_ip_j}$ and
$\Spbb{p_ip_j}$) and constants such as masses and coupling constants.

\begin{figure}[ht]
\begin{center}
\includegraphics{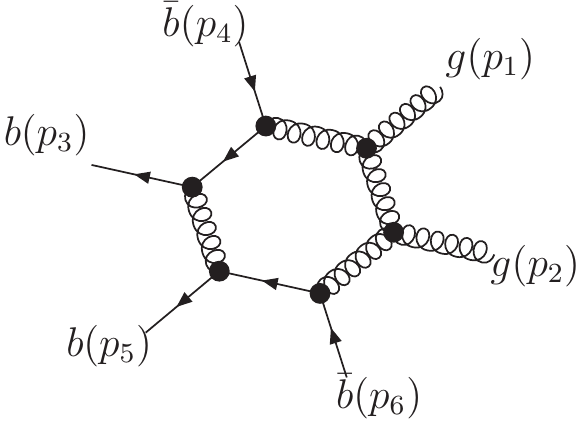}
\end{center}
\caption{A Feynman diagram contributing to the process
$gg\rightarrow b\bar{b}b\bar{b}$ at next-to-leading order
in \ac{QCD}.}
\label{fig:ggbbbb-d1498h23l1}
\end{figure}
The integrals in Equation~\eqref{eq:fpintegral} --- or rather form factors
that consist of combinations of these integrals --- are evaluated
through a \FortranXC{} library and translate directly to function
calls. However, for complicated $2\rightarrow4$ processes the number
of terms and the complexity of the coefficients in front of the form factors
can grow very large. For example, in the process
$gg\rightarrow b\bar{b}b\bar{b}$
we observe for the textual representation of the expression for
a single six-point diagram (see Figure~\ref{fig:ggbbbb-d1498h23l1})
a size of~9\mega\byte{} (43{,}918~terms).
Expressions of this complexity cannot be compiled efficiently by
standard \FortranXC{} compilers; often a compilation is not possible
at all, prohibited by the sheer size of the expression.
Some of the reasons for that are:
\begin{itemize}
\item typically, compilers implement algorithms that produce optimal
   code for relatively small subprograms at the expense of a
	non-linear time and/or resource consumption at compile time.
	The GNU compiler collection\footnote{version 4.3}, for example, uses a
	graph colouring algorithm for register allocation which is known
	to have a runtime that scales quadratically with the number of
	temporary variables.
	These algorithms fail for very large subprograms.
\item most compilers make no or little assumptions about the algebraic
   properties of binary operators. Especially for overloaded operators,
   properties such as commutativity or associativity cannot be specified.
	Therefore, expressions very often have to be evaluated unoptimised.
\item side-effects, aliasing and the possible dependence on global variables
   limit a compiler's possibilities of reusing the outcome of a function
	call\footnote{Overloaded operators have to be viewed as
	function calls, too.}.
	Therefore, \ac{CSE} does typically not include function calls.
\end{itemize}
Although this list of reasons might look to the reader like a list of
shortcomings of current compiler technology, there are good reasons
not to touch the current behaviour of the compilers:
for example, when overloading the operator `\verb|*|' with a matrix product,
commutativity should not be assumed by the compiler, and the result of
a function that returns random numbers or the current time should not
be~reused.

One alternative would be an extension of the target language by an appropriate
set of keywords for marking functions and operators as symmetric or free of
side-effects. \FortranXC, for example, has a keyword `\texttt{pure}' which
allows to specify a function as side-effect free.
However, further restrictions would be necessary to allow for fully automated
optimisation --- on the other hand these restrictions (e.g. absence of
pointers and global variables) can become prohibitive in other parts of
the program.

Another alternative is to put these additional assumption into a preprocessor
that then presents the already optimised source code to an existing
compiler. Within the \GOLEM{} implementation, we have chosen this solution,
which lead to the development of the program \haggies{}.
Although rooted in the field of high energy physics,
the possible applications of \haggies{} are much broader as will be
shown in the demonstration programs.

The remainder of this article is structured as follows:
Section~\ref{sec:algorithm} describes the algorithm used to transform
the input expressions. The installation and the system requirements
are briefly described in Section~\ref{sec:install}, then follows
a number of examples in Section~\ref{sec:examples}.
In the appendix we give a complete reference over
the language of the configuration and template files.

\section{Description of the Algorithm}\label{sec:algorithm}

\subsection{Overview}\label{ssec:algorithm:overview}
\begin{figure}[ht]
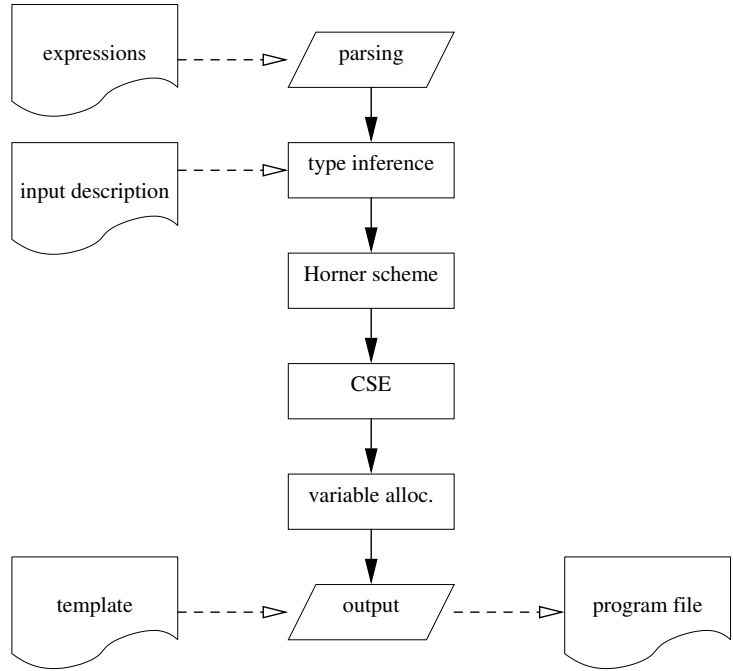

\begin{center}
\xfigbox{0.7\textwidth}{flowchart.pdftex}
\end{center}
\caption{Structure of the program \haggies{}. Dashed lines denote
data flow, continuous lines stand for control flow.}
\label{fig:flowchart}
\end{figure}

The program consists of several consecutive parts each of which runs
on the representation of the expression produced by the previous part.
A schematic overview is given in Figure~\ref{fig:flowchart}. Each of the
steps is described below.

\subsection{Parsing and \acs{AST}-Representation}\label{ssec:algorithm:parser}
The parsing of the textual representation of the input expressions
has the only challenge that an efficient memory representation has to
be used. For the parsing itself the Java parser generator \texttt{JavaCC}
is used, which produces a recursive descent parser~\cite{aho-sethi-ullman}
from an \ac{EBNF} description of the
expression grammar. At this point the program offers two interfaces:
the expression syntax used by
\FORM{}~\cite{Vermaseren:2000nd,Vermaseren:2006ag}
 which is fairly compatible
with the expression syntax of most computer algebra systems, and
a reader for Mathematica expressions, which distinguishes itself from
most other programs by the use of square brackets in function calls
instead of round brackets.

\begin{figure}[hp]
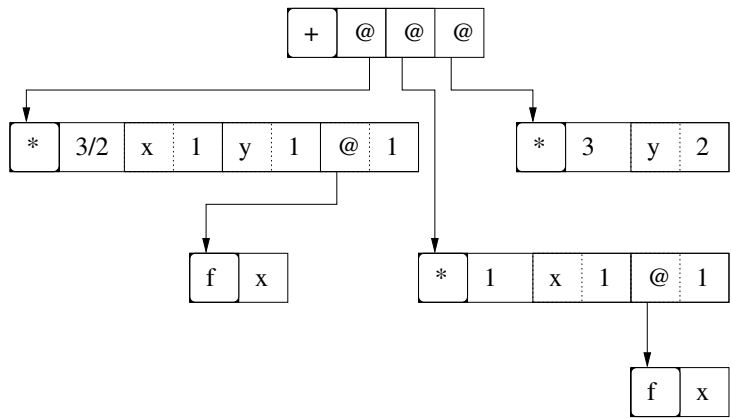

\begin{center}
\xfigbox{0.7\textwidth}{ast.pdftex}
\end{center}
\caption{\acs{AST} representation of the expression
$5/2xyf(x)-xf(x)\cdot(y-1)+3y^2$; here, the second term is already
expanded.
The dotted lines separate
basis and exponents inside a product. A mixed tree/array representation
is used to guarantee quick access to the elements.
The character `@' is used here to denote a~pointer.}
\label{fig:ast}
\end{figure}
The \ac{AST} implements commutativity and distributivity
to store a sorted, fully expanded representation of each expression.
Both sums and products use arrays to store their terms (resp. factors).
Figure~\ref{fig:ast} shows a pictogram of the memory representation of
a simple expression.

During the construction of the \ac{AST}
the expression is brought into a canonical form by expanding all products
and by sorting all factors (resp. terms)
inside products (resp.~sums).\footnote{Automatic expansion can be
suppressed by specifying the \texttt{-E} option at the command line.}
At this point we do not yet introduce a \ac{DAG} representation, as this is
achieved later during the \ac{CSE}.

\subsection{Type Inference}\label{ssec:algorithm:inference}
\lstset{language=c}
Although for many problems it is sufficient to work with a
single numerical data type throughout the whole calculation, it often
adds more flexibility to mix different data types.
As an example the user is referred to Section~\ref{ssec:examples:rk},
which implements a Runge-Kutta integrator for the differential
equation $\vec{y}^\prime=f(x,\vec{y})$. This example requires the
notion of at least two data types, a floating point type
(e.g. \lstinline!double!) for $x$ and a vector type (\lstinline!double[]!)
for $\vec{y}$ and $\vec{y}^\prime$. Without this flexibility the only
way out is to implement the algorithm for a fixed number of variables
$y_1, \ldots, y_n$ and to regenerate the code whenever $n$ needs to
be changed.

In the case of the \GOLEM{} implementation, having multiple data types
is a necessity rather than added flexibility. One has to deal with
integer indices, Laurent expansions in $(1/\varepsilon)$,
Taylor expansions in $\varepsilon$ and complex and real numbers
at the same time (see Table~\ref{tbl:golemtypes}).
\lstset{language=haggies}
\begin{table}[ht]
\begin{center}
\begin{tabular}{r@{\hspace{2em}}l}
\texttt{L} & \textit{Laurent expansion, such as
	$A/\varepsilon^2+B/\varepsilon+C+\mathcal{O}(\varepsilon)$}\\
\texttt{E} & \textit{Taylor expansion, such as
	$X + Y\varepsilon+Z\varepsilon^2+\mathcal{O}(\varepsilon^3)$}\\
\texttt{I} & \textit{Integer numbers}\\
\texttt{R} & \textit{Real numbers}\\
\texttt{C} & \textit{Complex numbers}
\end{tabular}
\end{center}
\caption{Basic types used for the objects in \GOLEM{} expressions.}
\label{tbl:golemtypes}
\end{table}
Between the objects of different types there are valid operations
such as $\mathtt{C}\cdot\mathtt{C}\rightarrow\mathtt{C}$ or
$\mathtt{L}\cdot\mathtt{E}\rightarrow\mathtt{L}$. On the other hand,
ill-defined expressions, such as 
$\mathtt{C}/\mathtt{L}$, indicate an error in an earlier step of the
calculation and should be rejected. The type system of \haggies{}
allows also for function types and implicit coercions.
Table~\ref{tbl:golem-definitions} shows an extract of the
list of definitions for the \GOLEM{} type system.

The program \haggies{} expects each symbol that appears
on the right-hand side of a calculation to be defined with
a type. The type of each subexpression is inferred starting
at the leaves of the \ac{AST} using the type information
of the operations and functions and if necessary, by inserting
coercions. If for any subexpression the type cannot be inferred
by the defined rules the program reports an error and terminates.
It is, however, not considered an error if there is more than one
possibility to infer the type of a subexpression;
the program will choose one possibility, relying on the user
to define a sound type system in the configuration files\footnote{%
This also means that there is no backtracking if the inference 
fails beyond a choice~point.}.

\begin{table}[ht]
\begin{center}
\begin{tabular}{ll@{\hspace{2em}}l}
\textbf{operations}&&\\
$\mathtt{L}\cdot\mathtt{E}\rightarrow\mathtt{L}$ & &\\
$\mathtt{L}+\mathtt{L}\rightarrow\mathtt{L}$ & &\\
$\mathtt{I}/\mathtt{I}\rightarrow\mathtt{R}$ & &\\
\ldots &&\\
\textbf{coercions}&&\\
$\mathtt{I}\rightarrow\mathtt{R}$&&\\
$\mathtt{R}\rightarrow\mathtt{C}$&&\\
$\mathtt{C}\rightarrow\mathtt{L}$ &&
	\textit{Laurent expansion with $A=B=0$.}\\
\ldots &&\\
\textbf{definitions}&&\\
$\varepsilon$:&$\mathtt{E}$ &\\
$m_T,m_W,e,g_s,\ldots$:&$\mathtt{R}$ &
	\textit{masses and coupling constants}\\
$\Spaa{p_ip_j},\Spbb{p_ip_j}$:&$\mathtt{C}$ &
\textit{spinor products}\\
$I_3^n(l_1, l_2)$:&$\mathtt{I},\mathtt{I}\rightarrow\mathtt{L}$ &\\ 
\ldots &&
\end{tabular}
\end{center}
\caption{Extract of the \GOLEM{} type system.}
\label{tbl:golem-definitions}
\end{table}

The inferred types will become relevant in
Section~\ref{ssec:algorithm:cse}, where \ac{CSE} is
discussed: consider the expression $f(xy)+g(xy)$
and the following, generated program fragment:
\begin{lstlisting}[language=pascal]
	t1 := x*y;
	t2 := f(t1);
	t3 := g(t1);
	t3 := t2+t3;
\end{lstlisting}
The newly introduced variable \verb|t1| can have a type different
from that of \verb|t2| and \verb|t3|. In a strongly typed language
this type information has to be made available to prepend the above
code segment by the necessary declarations, such as the following:
\begin{lstlisting}[language=pascal]
	var t1: integer;
	var t2: real;
	var t3: real;
\end{lstlisting}

Another reason for attaching type information to all
subexpression is the use of user-defined data types
such as multiprecision numbers or to allow the
use of interval arithmetic\footnote{See also the examples
in Section~\ref{ssec:examples:rk}.}; in languages without operator
overloading these cases require to replace the operator expressions
by function or method calls.

\subsection{Multivariate Horner Scheme}\label{ssec:algorithm:horner}
The first transformation applied to the expressions is a
multivariate Horner scheme.
This step basically follows the method proposed
in~\cite{Ceberio:03}. For univariate polynomials the Horner scheme
provides the fastest implementation of evaluating the polynomial at a
given position, i.e. the evaluation can be performed with the least number
of multiplications and additions~\cite{Knuth:TAOCP1}.
In the multivariate case, however,
it is not clear a priori which evaluation scheme leads to the smallest number
of arithmetic operations. Consider, for example the polynomial
\begin{equation}
f(x_1,x_2,x_3)=x_1^3x_2+x_1^2x_3+x_1^2x_2x_3\text{.}
\end{equation}
Obviously, extracting $x_1^2$ from each term saves the largest number
of operations, and we obtain
\begin{equation}
f(x_1,x_2,x_3)=x_1^2(x_1x_2+x_3+x_2x_3)\text{.}
\end{equation}
The next decision already becomes less trivial, as we can either select
$x_2$ or $x_3$, and one obtains
\begin{subequations}
\begin{align}
f(x_1,x_2,x_3)&=x_1^2(x_2(x_1+x_3)+x_3)\quad\text{or}%
\label{eq:factorization:a}\\
f(x_1,x_2,x_3)&=x_1^2(x_1x_2+x_3(1+x_2))%
\label{eq:factorization:b}
\end{align}
\end{subequations}
respectively. The representation in Equation~\eqref{eq:factorization:a}
requires one multiplication less than Equation~\eqref{eq:factorization:b}.
This simple example already shows that the order
in which the variables are extracted
will impact the efficiency of the evaluation.

In general, a multivariate Horner scheme can be generated by
Algorithm~\ref{alg:mvh}. At each step one or more variables
$x_1^{m_1}\cdots x_n^{m_n}$ ($0\leq m_i$) are selected
and the polynomial is split into terms that contain the
selected product of variables and terms $a_0$ from which this
product cannot be factored out. The algorithm still leaves the heuristics
behind \texttt{select\_coeffs} undefined. A global optimisation would
try to minimise the number of multiplications of the final expression,
since the number of additions remains constant. On the other hand it is clear
that such a global strategy needs to consider far too many possibilities
and hence cannot be implemented efficiently.
\begin{algorithm}
\caption{$\mathtt{multivariate\_horner}(f)$}
\label{alg:mvh}
\begin{algorithmic}
\REQUIRE polynomial $f(x_1,\ldots,x_n)$
\STATE{$\vec{m}\leftarrow \mathtt{select\_coeffs}(f)$}
\STATE{Find $a_1$, $a_0$ such that
	$f(x_1,\ldots,x_n)=x_1^{m_1}\cdots x_n^{m_n}a_1+a_0$.}
\STATE{$b_1\leftarrow \mathtt{multivariate\_horner}(a_1)$}
\STATE{$b_0\leftarrow \mathtt{multivariate\_horner}(a_0)$}
\RETURN{$x_1^{m_1}\cdots x_n^{m_n} b_1 + b_0$}
\end{algorithmic}
\end{algorithm}

The simplest strategy that runs at each step of the algorithm
in linear time is to select only one variable at a time.
The authors of~\cite{Ceberio:03} suggest the selection of the
variable which appears in the highest number of terms, which leads
to the highest immediate decrease of arithmetic operations at a given
step. Our implementation allows in a very straight forward manner to
add new strategies but it proves difficult to come up with strategies
which, in combination with \ac{CSE}, outperform the original method.

\subsection{\texorpdfstring{\acl{CSE}}{Common Subexpression Elimination}}
\label{ssec:algorithm:cse}
\acl{CSE} denotes a source code transformation in which temporary variables
are introduced for each subexpression such that it is only calculated once
and can be reused at a later point in the calculation without having to be
calculated again.\footnote{See, for example~\cite{aho-sethi-ullman}.}
A great simplification for our program is the fact that
we transform only expressions and therefore can neglect all complications
arising from control structures such as loops and jumps.

We consider again the example from Figure~\ref{fig:ast}.
The Horner form for this expression as produced by our program is
\begin{equation}
x\cdot\left(\frac32yf(x)+f(x)\right)+3y^2\text{.}
\end{equation}
If $f(x)$ encodes a computationally expensive procedure call
a good programmer would have written\footnote{%
This program fragment assumes that the division operator denotes
a floating point division, not integer division.}
\begin{lstlisting}[language=pascal]
	t := f(x);
	result := x*(3/2*y*t+t) + 3*y*y;
\end{lstlisting}
\ac{CSE} does precisely the same; the program
produced from this example is shown below:
\begin{lstlisting}[language=pascal]
	$1 := f(x);
	$2 := $1 * y;
	$3 := 3/2 * $2;
	$4 := $1 + $3;
	$5 := y * y;
	$6 := 3 * $5;
	$7 := $4 * x;
	$8 := $6 + $7;
	result := $8;
\end{lstlisting}

Since one of the main goals is to minimise the number of multiplications
the program takes special care about high powers of variables.
For a single variable the most efficient way
to calculate an integer power is the binary exponentiation which uses the
binary representation of the exponent in order to compute the power with
the least number of multiplications. For example $x_3^5$ can be computed
as
\begin{equation}
x_3^5=x_3^{2^0+2^2}=x_3^{1+2\cdot 2}=x_3\cdot(x_3^2\cdot x_3^2)\text{,}
\end{equation}
which requires 3 multiplications, since the value of $x_3^2$ can be
stored and reused. If we apply this trick to each variable of
the term $x_1^3x_2^4x_3^5$ we can compute the expression
with nine multiplications. A more efficient way to compute the same
result is multi-exponentiation: here, we consider the binary representation
of a vector of exponents,
\begin{equation}
\begin{pmatrix}3\\4\\5\end{pmatrix}
=\begin{pmatrix}1\\0\\1\end{pmatrix}
+2\left(
\begin{pmatrix}1\\0\\0\end{pmatrix}
+2\begin{pmatrix}0\\1\\1\end{pmatrix}
\right)
\end{equation}
The according factorisation of the product would be
\begin{equation}
(x_1x_3)(x_1\cdot(x_2x_3)^2)^2
\end{equation}
which can be calculated with only six multiplications.
This can also be seen from the output produced by \haggies{}:
\begin{lstlisting}[language=pascal]
	$1 := x1*x3;
	$2 := x2*x3;
	$3 := $2*$2;
	$4 := $3*x1;
	$5 := $4*$4;
	$6 := $1*$5;
\end{lstlisting}

\subsection{Register Allocation}\label{ssec:algorithm:registers}
It would, of course, be very inefficient if the algorithm stopped
at this point since a new variable is introduced for each operation.
Many of the intermediate results need to be saved only for
a short period and can be discarded after a small number
of instructions. Hence in a further step \haggies{} needs to determine
the life span of each intermediate result and replace the virtual
variables \verb|$1|, \verb|$2|, \dots by a much smaller number of actual
local variables. At this point, an actual compiler would try to match
the local variables with physical processor registers and introduce
so-called memory spills whenever the number of registers on the target
machine is not sufficient. We can, however, assume an unlimited number
of registers and try to minimise the number of actually used registers;
the allocation of physical registers for these intermediates is left
to the~compiler.

Usually this step is implemented by a graph colouring
algorithm~\cite{Chaitin:81}:
the graph is constructed by connecting all virtual variables with an overlapping
life range. This graph is then coloured by the number of colours corresponding
to the number of available registers. If the graph is not colourable memory
spills are introduced. This type of algorithm typically produces allocations
close the optimal solution but the price to pay is a worst-case performance
which grows quadratically in the number of life ranges.
For very large expressions this approach can render compilation impossible.

The program \haggies{} implements a single pass strategy
that runs in linear time~\cite{Poletto:99}. This \textit{Linear
Scan Register Allocation} algorithm has been proposed with
Just-In-Time compilation in mind, where compile time and run time
performance are equally important. In our case its striking feature is
the linear scaling behaviour.

We run this algorithm separately for each different data type
as defined by the type inference. For the example from Figure~\ref{fig:ast}
it suffices to introduce two variables.
After register allocation the resulting program is as follows:
\begin{lstlisting}[language=pascal]
	t1 := f(x);
	t2 := t1 * y;
	t2 := 3/2 * t2;
	t1 := t1 + t2;
	t2 := y * y;
	t2 := 3 * t2;
	t1 := t1 * x;
	t1 := t1 + t2;
	result := t1;
\end{lstlisting}

\subsection{Output and Syntax Transformations}\label{ssec:algorithm:output}
As already mentioned, one important consideration in the design
of the program \haggies{} is its independence from the target language.
For a large class of problems and languages only minor differences
distinguish the output, such as different notations for the assignment
operator (\verb|=| as opposed to \verb|:=|), the presence of a semicolon
at the end of a statement or requirements with respect to indentation
and line continuations
in fixed-form and some scripting languages. These differences are very
often solved by running  a standard text processor, such as \texttt{sed}
over the output produced by some computer algebra program.
However, some output formats can not easily be achieved by simple text
transformation. Some common problems are listed below:
\begin{itemize}
\item The (non-)existence of an exponentiation operator:
In \texttt{C} this operator has to be emulated by a function
call to the `\texttt{pow}' function, some languages use the
operator `\verb|**|' instead of `\verb|^|'. In some languages
the function name depends on the data type, e.g.
`\texttt{powf}' and `\texttt{powd}' for single resp. double precision
numbers.
\item Languages such as \texttt{Lisp} require the expression in
prefix notation and/or use a different notation for function
calls (`\texttt{(f x)}' instead of `\texttt{f(x)}').
\item The division operator between two integer numbers
in some languages is interpreted as an integer division.
Hence, a naive translation of $1/3$ yields $0$ instead of
the correct numerical value $0{.}333\ldots$.
\item The use of non-standard data types in languages without
operator overloading needs to be replaced by function or
method calls. In \texttt{Java}, the use of the class \texttt{BigDecimal}
requires method calls such as `\verb|x.add(y)|' rather than `\verb|x+y|'
similar problems arise in the context of multiprecision libraries
in other languages.
\end{itemize}

We solve the above problems by allowing for the specification of
patterns for each operation along with each declared data type.
The declaration of a data type represented by the \texttt{BigDecimal}
class in \texttt{Java} could be declared as follows:

\lstset{escapeinside={<+}{+>}}
\begin{lstlisting}[language=haggies,numbers=left,%
   numberstyle={\footnotesize}]
@type
   R = "BigDecimal", "%s.add(%s)",<+\label{lst:ex01:LINE2}+>
      "%s.subtract(%s)", "%s.negate()";<+\label{lst:ex01:LINE3}+>
   I = "long";<+\label{lst:ex01:LINE4}+>
   F = "double";
@operation
   R * R -> R =<+\label{lst:ex01:LINE7}+>
      "%s.multiply(%s)", "%s.divide(%s)";
   R * I -> R = "%s.multiply(%s)", "%s.divide(%s)",
      "BigDecimal.valueOf(%2$s).divide(%1$s)";<+\label{lst:ex01:LINE10}+>
\end{lstlisting}
\lstset{language=haggies}
Lines \ref{lst:ex01:LINE2}--\ref{lst:ex01:LINE3}
define the type `\lstinline!R!' which in \texttt{Java}
is represented by the class \texttt{BigDecimal}; the strings that
follow the type name are, in this order, the pattern to be applied
for an addition, subtraction and the unary minus. The patterns
are defined by the syntax for the \texttt{Java} class
\texttt{java.text.Formatter}~\cite{java:Formatter}.
Similarly, line~\ref{lst:ex01:LINE4}
defines a type `\lstinline!I!' represented by
a long integer in \texttt{Java}. In this case we can leave out the
patterns since the default operators can be used.
Lines~\ref{lst:ex01:LINE7}--\ref{lst:ex01:LINE10}
define the multiplication and division of these two data types.
Here, the first two patterns on the right-hand side define multiplication
and division. An optional third pattern denotes the reverse division
(\lstinline!I/R! rather than \lstinline!R/I!).

Similar transformations can be defined for symbols and functions. The
examples below collect some commonly used patterns. A systematic description
of the syntax can be found in Appendix~\ref{ssec:reference:in}. The character
\lstinline!" "! denotes a blank character.
\begin{lstlisting}[language=haggies]
@define
   sin, cos : F -> F = "Math.%s(%s)";
   mat : I, I -> F = "m[%2$s][%3$s]";
   vec... : I -> F = "%s[%s]";
   abs : R -> R = "%2$s.abs()";
   ARRAY3 : R, R, R = "{%2$s, %3$s, %4$s}";
\end{lstlisting}
\lstset{language=haggies}
The first line contains names which need to be qualified by
a module or class name. The function `\lstinline!mat!' is a
placeholder for the array `\lstinline!m!'. The next line contains
an ellipse on the left hand side of the definition which applies
to all symbols that start with the prefix `\lstinline!vec!'.
The function `\lstinline!abs!' is transformed into the according
method call and the last example shows a function which is transformed
into an array literal.

These examples also show a couple of shortcomings and pitfalls.
First of all, there is no function overloading in \haggies{}.
It is, for example, not allowed to define the function `\lstinline!abs!'
twice with a different type. Currently, there is also no notion of
argument lists of variable length. Hence, we have defined `\lstinline!ARRAY3!'
in the above example and would have to define different symbols
for arrays of different size. The user should be aware that the pattern
\lstinline|"%1$s"| holds the function name and therefore the arguments start
at \lstinline|"%2$s"|.

\clearpage
\section{Installation and Requirements}\label{sec:install}
\subsection{Requirements}
\haggies{} runs on a \texttt{Java} virtual machine
that supports the \texttt{Java} API in version 1{.}5 or higher.
For compilation of the sources from scratch also \texttt{JavaCC}
version 5{.}0 or higher is needed.\footnote{%
\texttt{JavaCC} is available from \url{https://javacc.dev.java.net}.}
The examples have additional
requirements which are described separately with each example.
The compilation is based on Apache's Ant tool.

\subsection{Download}
All files can be downloaded from the URL
\begin{center}
\texttt{http://www.nikhef.nl/\~{}thomasr/filetransfer.php?file=\it{filename}}
\end{center}
where \textit{filename} is one of the following:
\begin{description}
\item[\texttt{haggies.jar}] containing the precompiled classes as
\ac{JAR} file. This is the only file required to run \haggies{}
\item[\texttt{haggies-src.jar}] containing the source code of the project
including the demo programs. It should be noted, that \texttt{JavaCC}
version~4{.}3 or newer and Apache Ant are required in order to compile
the project.
\item[\texttt{haggies-demo.zip}] containing the demo files only.
\end{description}

\subsection{Compilation}
For those users who prefer to compile the project from the source
files we describe here the installation procedure. This section
can be skipped when working with the precompiled \acs{JAR} file.

In the following we assume a \texttt{Linux} like system with
the \ac{JDK} installed.
\begin{enumerate}
\item The sources need to be extracted into a new directory:
   copy or move the file \texttt{haggies-src.jar} to that
	directory and change the working path to the target directory.
	Run \texttt{jar~xf~haggies-src.jar}.
\item The directory now contains, amongst others, the file
   \begin{verbatim}
   site-properties.template
   \end{verbatim}
   which needs to be copied
	to \texttt{site.properties}. Open the file in an editor
	and change the directory \texttt{javacchome} to the
	installation path of \texttt{JavaCC}.
\item Run \texttt{ant} to compile everything. If the compilation
   was successful the file \texttt{dist/haggies.jar} has been
	created.
\item With the GNU \texttt{Java} compiler installed there is
   the optional choice to create a binary executable from the same
	sources. To do this, in \texttt{site.properties} the option
	\texttt{gcj.installed} must be set to `\texttt{true}';
	then one can run \texttt{ant~bin}. The executable is 
	written to \texttt{dist/haggies}.
\item The files in the subdirectory \texttt{build} are not needed
   anymore and can be removed with the command `\texttt{ant~clean}'.
\end{enumerate}

\section{Examples}\label{sec:examples}
In order to use the example files one needs to download and unpack
the file \texttt{haggies-demo.zip}.\footnote{This file is not needed if
\texttt{haggies-src.jar} has been downloaded already.}
The examples are located in the subdirectory~\texttt{demo}.

\subsection{Runge-Kutta Integrator}\label{ssec:examples:rk}
\begin{tabular}{lp{0.6\textwidth}}
\bf Synopsis: & comparison of different Runge-Kutta methods
                applied to the Lorenz oscillator \\
\bf Objective: & basic usage of \haggies{},
                 generating output for different target languages \\
\bf Requirements: & \FORM{}, \FortranXC{} compiler \\
\bf Directory: & \texttt{demo/rk}
\end{tabular}

\subsubsection{Overview}
In this example, we construct implementations of the Runge-Kutta
method for different programming languages from the according
coefficients. Given the initial value problem
\begin{equation}
	\frac{\mathrm{d}}{\mathrm{d}x}\vec{y}(x)=f(x,\vec{y}),
	\quad
	\vec{y}(x_0)=\vec{y}_0\text{.}
\end{equation}
The series
\begin{align}
\vec{y}_{n+1}&=\vec{y}_n+h\sum_{i=1}^s b_i\vec{k}_i,\quad
\sum_{i=1}^sb_i=1\\
\intertext{with}
\vec{k}_i&=f(x_n+c_ih,\vec{y}_n + a_{s1}h\vec{k}_1+\ldots+a_{i,i-1}hk_{i-1})
\end{align}
defines an explicit Runge-Kutta method.
The choice of the coefficients $b_i$, $c_i$ and $a_{ij}$ determines
the order~$p$,
\begin{equation}
\vec{y}(x_0+nh)=\vec{y}_n+\mathcal{O}(h^p)\text{.}
\end{equation}
In practice, one works with two different sets $b_i$ and $d_i$
which are chosen such that replacing $b_i$ by $d_i$ in the above
equations results in a method of order $(p-1)$. The difference
between the two results gives an estimate for the error and allows
for an adaptive choice of $h$.

As an application of the Runge-Kutta method we integrate
the Lorenz oscillator
\begin{equation}\label{eq:lorenz-osc}
\frac{\mathrm{d}}{\mathrm{d}t}\begin{pmatrix}
x\\y\\z
\end{pmatrix}=\begin{pmatrix}
\sigma\cdot(y-x)\\
x\cdot(\rho-z)-y\\
x\cdot y-\beta\cdot z
\end{pmatrix}
\end{equation}
for $\sigma=10$, $\beta=8/3$, $\rho=28$ and $0\leq t\leq20$.
This set of coupled, non-linear differential equations describes
an unstable dynamics which is very sensitive to numerical inaccuracies
and therefore provides a good benchmark for different integration methods.

\subsubsection{\texttt{C}-Implementation}
The file \texttt{rk.frm} encodes the coefficients for different
methods~\cite{Cash:90,Bogacki1989321,Dormand198019,Fehlberg70}
as a \FORM{} program which generates
the expressions for $\vec{y}_{n+1}$ and
$\vec{z}_{n+1}\equiv\vec{y}_{n+1}\vert_{b_i\rightarrow d_i}$.
Although there are more direct methods for constructing an implementation
of the integrator, this example shows how one can produce different
programs from the same computer algebra source using \haggies{} as
a converter. For the simplest implemented integrator of order~$2$
the \FORM{} program generates the two expressions
\begin{subequations}\label{eq:heun}
\begin{align}
y_1&=y_0 + 1/2\cdot f(x_0 + h,y_0 + f(x_0,y_0)\cdot h)\cdot h
+ 1/2\cdot f(x_0,y_0)\cdot h;\\
z_1&= y_0 + f(x_0,y_0)\cdot h;
\end{align}
\end{subequations}
For the higher order methods these expressions are substantially longer.

The next step is to generate a \texttt{C}-file from the above expressions
using \haggies{}. The program requires a configuration file
(Listing~\ref{lst:rk/c.in})
and a template file (Listing~\ref{lst:rk/c.out})
which together determine the transformations.

The configuration file defines all occurring symbols, the
admitted operations between them and their representation
in the output file. The template file, on the other hand,
determines the structure of the output file and the format
of the statements and declarations.

\begin{lstlisting}[float,language=haggies,frame=tb,%
	caption={[Runge-Kutta Example: \texttt{c{.}in}]%
	Configuration file `\texttt{c{.}in}' %
	to produce \texttt{C} output for the %
	Runge-Kutta example. %
	The type `\lstinline!T!' could in principle be chosen to be a vector type.},%
	captionpos=b,%
	label={lst:rk/c.in},%
	firstnumber=1,%
	numbers=left,stepnumber=5,numberfirstline=false,%
	numberstyle={\footnotesize}]
@language form -> c;<+\label{lst:rk/c.in:1}+>
@type<+\label{lst:rk/c.in:type(}+>
	T = "float";
	S = "float";<+\label{lst:rk/c.in:type)}+>
@coerce<+\label{lst:rk/c.in:coerce(}+>
	@int -> S = "%s.0";
	@int/@int -> S = "%s.0/%s.0";<+\label{lst:rk/c.in:coerce)}+>
@define<+\label{lst:rk/c.in:define(}+>
	x0, h : S;
	y0, y1, z1: T;
	f : S, T -> T;<+\label{lst:rk/c.in:define)}+>
@operator<+\label{lst:rk/c.in:operator(}+>
	S * S -> S;
	S * T -> T;<+\label{lst:rk/c.in:operator)}+>
@polynomial x0, y0;<+\label{lst:rk/c.in:poly}+>
\end{lstlisting}

\lstset{language=haggies}
The \lstinline!@language! statement in line~\ref{lst:rk/c.in:1}
of Listing~\ref{lst:rk/c.in} defines
the languages, both of the input and the output file.\footnote{
See also Appendix~\ref{r:language}}
In line \ref{lst:rk/c.in:type(}--\ref{lst:rk/c.in:type)}
we define the two different data types that appear in the problem,
a scalar numeric type `\lstinline!S!' for $x_0$ and $h$ and
a vectorial type `\lstinline!T!' for $\vec{y}_0,\vec{y}_1,\ldots$.
Here, we have assigned the \texttt{C} data type \lstset{language=c}
\lstinline!float! to both formal data types which corresponds to
the case where $\vec{y}$ is just one-dimensional.

\lstset{language=haggies}
In Lines~\ref{lst:rk/c.in:coerce(}--\ref{lst:rk/c.in:coerce)} we
define coercions, which are applied whenever necessary. The data types
\lstinline!@int! and \lstinline!@int/@int! correspond to the data types
of integer and rational literals. If the program encounters the expression
$5\vec{y}$ then first the coercion \lstinline!@int->T! is applied
after which the operation \lstinline!S*T->T! can be matched.
The optional right-hand side of a \lstinline!@coerce! statement defines
the textual transformation
of the current expression, hence the literal `\lstinline!5!' will be
transformed into `\lstinline!5.0!'. In the case of the data type
\lstinline!@int/@int! the two fields of the pattern correspond to
numerator and denominator of the fraction.\footnote{
See also Appendix~\ref{r:coerce}}

The \lstinline!@define! statement
(Lines~\ref{lst:rk/c.in:define(}--\ref{lst:rk/c.in:define)})
declares all symbols which are valid in the input expression
and associates them with a type. In order to denote groups of
symbols with a common prefix and/or suffix one can use the ellipsis;
instead of `\lstinline!y0, y1: T!' we could also have written
`\lstinline!y... : T!'. The data type of \lstinline!f! is a
functional type which maps the argument list of type
`(\lstinline!S!, \lstinline!T!)' to the return type `\lstinline!T!'.%
\footnote{See also Appendix~\ref{r:define}}

So far we have not defined yet which operations are valid between
the different data types. In the \lstinline!@operator! section
in Lines~\ref{lst:rk/c.in:operator(}--\ref{lst:rk/c.in:operator)}
we define the multiplication\footnote{and therefore also division}
between the data types `\lstinline!S!' and `\lstinline!T!'.
An implicit declaration of addition, subtraction
and negation is always made together with the \lstinline!@type! statement.
Hence, also the operations `\lstinline!T+T->T!' and `\lstinline!S+S->S!
are permitted. A detailed reference on these two statements can be found
in Appendices~\ref{r:operator} and~\ref{r:type}.

The last statement of the configuration file forms the
\lstinline!@polynomial! statement in Line~\ref{lst:rk/c.in:poly}.
Here one lists all symbols that should be considered as parts of the
monomials when applying the Horner scheme to the expression. All
remaining symbols form the coefficients of the monomials together with
the numerical coefficients. The keyword \lstinline!@polynomial! can also
be abbreviated by \lstinline!@poly!.\footnote{%
See also Appendix~\ref{r:polynomial}}

\begin{lstlisting}[float,language={[c]haggiest},frame=tb,%
	caption={[Runge-Kutta Example: \texttt{c{.}out}]%
	Template file `\texttt{c{.}out}' %
	to produce \texttt{C} output for the %
	Runge-Kutta example. The markup which is processed by \haggies{}
	is marked in red.},%
	captionpos=b,%
	label={lst:rk/c.out},%
	firstnumber=1,%
	numbers=left,stepnumber=5,numberfirstline=false,%
	numberstyle={\footnotesize}]
float integrate(
      float (*f)(float, float),
      float x0,
      float y0,
      float h,
      float& err)
{<+\label{lst:rk/c.out:LOOP1(}+>[%
   @for symbols match="\\$(\\d+)" format="t%s" %]
   [% type.repr %] [% $_ %];[%
   @end @for symbols %]<+\label{lst:rk/c.out:LOOP1)}+>
{<+\label{lst:rk/c.out:LOOP2(}+>[% @for instructions %]
   [% @select $_ match="(.).*" format= "%s"
      @case $ %][% $_ match="\\$(\\d+)"
         format="t%s"%][%
      @else %][% $_ %][%
      @end @select $_
   %] = [% expression match="\\$(\\d+)"
      format="t%s"%];[%
   @end @for instructions %]<+\label{lst:rk/c.out:LOOP2)}+>
   err = abs(y1 - z1);
   return y1;
}
\end{lstlisting}

The program \haggies{} transforms the expressions from the input file
into a sequence of assignments that calculates this expression numerically.
In general, these statements need to be embedded syntactically in some
kind of module, class or procedure.
Since in our example the \texttt{C}-output is for demonstration purposes
only the template file in Listing~\ref{lst:rk/c.out} has been kept as
simple as possible. As we progress through the further examples
we will encounter more complex template files.
\lstset{language={[c]haggiest}}
Template files can contain parts of a program in the target language,
which will be emitted to the output file verbatim,
plus markup tags which are enclosed by a pair of brackets
\lstinline![% ... %]!, which are highlighted in colour througout this article.
Inside a tag a single quote starts a comment that spans to the end of
the tag.

In Line~\ref{lst:rk/c.out:LOOP1(} we start a loop which
iterates over all symbols that are introduced by the \ac{CSE}; these
symbols have the names `\lstinline!$1!', `\lstinline!$2!' and so on.
A loop has the general form
\begin{lstlisting}[language={[c]haggiest}]
[% @for <+$name$+> <+$opt_1$[+>=<+$value_1$] %
      +> <+$opt_2$[+>=<+$value_2$]+> <+$\ldots$+>
      %]loop body[%
   @end @for %]
\end{lstlisting}
The $name$ of the iterator in our case is `\lstinline!symbols!'; inside
the loop it defines the name of each symbol \lstinline![% $_ %]!
together with its type name \lstinline![% type.name %]!
(here `\lstinline!S!' and `\lstinline!T!') and its type
representation \lstinline![% type.repr %]!
(here `\lstinline!float!' in both cases). In general the
type name is the left hand side of a type declaration and
the type representation the right hand side; the type representation
defaults to the type name if the right hand side was omitted in the
configuration file. Most loops also define the two Boolean values
\lstinline![% is_first %]! and \lstinline![% is_last %]!
which can be used inside an \lstinline![% @if %]! statement
to output some section only before the first or after the last
element of a loop\footnote{The difference to placing the code just before
or after the loop becomes important for empty iterators which trigger the
loop zero times.}.
The loop is terminated in Line~\ref{lst:rk/c.out:LOOP1)}.
Note that all arguments of the \lstinline![% @end @for %]!
statement are ignored and merely serve as comments which makes
it easier to maintain nested loops.

The pair of options `\lstinline!match!' and `\lstinline!format!' is
common to most tags and allows a pattern based transformation of the
symbol concerned. The value supplied to the option `\lstinline!match!'
is interpreted as a regular expression according to the syntax
accepted by the \texttt{Java} class
\texttt{java.util.regex.Pattern}~\cite{java:Pattern};
the value of `\lstinline!format!' is a \texttt{Java}
\texttt{printf}\footnote{The format is defined by the methods of the
class \texttt{java.util.Formatter}. Because of the similarity to the
\texttt{printf} format in \texttt{C} we call it \texttt{Java}
\texttt{printf}~format.} format~\cite{java:Formatter}.
The groups\footnote{A \emph{group} denotes a subpattern delimited by
a pair of parentheses. Groups are used in substitutions to use
parts of the match in the replacement.} matched by the regular expression
are taken as the fields of the format. In our example, we have the pair
\lstinline![% match="\\$(\\d+)" format="t%s" %]!;
if applied to the symbol `\$123' the matched group is the substring `123'
which is substituted for the field `\lstinline!%s!' in the format
and hence transforms to the symbol `t123'.

The second loop of the program spans the
Lines~\ref{lst:rk/c.out:LOOP2(}--\ref{lst:rk/c.out:LOOP2)}.
Here we iterate over the instructions that compute the expressions.
In this loop, the symbol \lstinline![% $_ %]!
denotes the left hand side of the assignment.
Furthermore, one can access the right hand side expression through
the tag \lstinline![% expression %]! and its type through
\lstinline![% type.name %]!.
It should be mentioned that for the right hand side expression
the match-format pair is not applied to the textual representation
of the expression but to each single symbol that occurs inside the
expression and is simply ignored for symbols that do not match the
regular expression.

On the left hand side of the assignment we have put a select
statement. This is necessary because the match-format transformation
should only be applied to symbols starting with a `\$' but not to
`\lstinline!y1!' and `\lstinline!z1!'. The regular expression
\lstinline!"(.).*"! picks only the first letter of the symbol which
is then compared to the values in the \lstinline![% @case %]!
clauses. The general form of a select statement is
\begin{lstlisting}[language={[c]haggiest}]
[% @select <+$name$+> <+$opt_1$[+>=<+$value_1$]+> <+$opt_2$[+>=<+$value_2$]%
  +> <+$\ldots$+>
   @case <+$value_{11}$+> <+$value_{12}$+> <+$\ldots$+>
      %]branch 1[%
   @case <+$value_{21}$+> <+$value_{22}$+> <+$\ldots$+>
      %]branch 2[%
	   <+$\vdots$+>
   @else
      %]branch n[%
   @end @select %]
\end{lstlisting}
The else-branch can be omitted.

The \texttt{C}-file is generated by running \haggies{}
on the input using the above files  as configuration and
template files.
In the simplest case the command looks as follows
\begin{verbatim}
java -jar haggies.jar -cc.in -tc.out -omethod_h.c method_h.txt
\end{verbatim}
Here we assume that the Equations~\eqref{eq:heun} are found in
the file `\texttt{method\_h.txt}', which must define the symbols
`\texttt{y1}' and `\texttt{z1}' by definitions of the form
$\langle\text{symbol}\rangle$\;`\texttt{=}'\;$\langle\text{expression}\rangle$%
\;`\texttt{;}'.

The option \texttt{-c} specifies
the configuration file, \texttt{-t} specifies the template and
\texttt{-o} the output file. If the option \texttt{-o} is omitted
the output is written to the standard output; if no input file is given
the input is read from the standard input. If we add the option
\texttt{-V2} we will find the following output on the screen\footnote{
A full list of options can be obtained if the program is invoked
with the argument \texttt{--help}. See also Appendix~\ref{ssec:reference:cli}}:

\begin{verbatim}
          y1:       5 (*),       4 (+),       0 (.),       3 ()
          z1:       1 (*),       1 (+),       0 (.),       1 ()
 (optimised):       4 (*),       4 (+),       0 (.),       2 ()
\end{verbatim}
This statistics gives information about the number of operations
in each of the unoptimised expressions (here \texttt{y1} and \texttt{z1})
and the cost of the program in the output. From left to right the numbers
denote the number of multiplications/divisions, the number of
additions/subtractions, the number of dot products and the number of
function calls. As expected for the Heun method, the program requires
two function calls; since we start from very simple input expressions
there are no big savings in the number of multiplications.
In more complicated examples we will find much larger ratios
between input and output in the first column.
The generated \texttt{C}-code can be found in Listing~\ref{lst:rk/method_h.c}.
\begin{lstlisting}[float,language={c},frame=tb,%
	caption={[Runge-Kutta Example: \texttt{method_h{.}c}]%
	The output produced from the template file `\texttt{c{.}out}' %
	when applied to the expressions of the Heun method},%
	captionpos=b,%
	label={lst:rk/method_h.c},%
	firstnumber=1]
float integrate(
      float (*f)(float, float),
      float x0,
      float y0,
      float h,
      float& err) {
   float t1;
   float t2;
   float t3;
   float t4;

   t4 = h+x0;
   t1 = f(x0,y0);
   t1 = h*t1;
   t2 = t1+y0;
   t3 = f(t4,t2);
   t3 = h*t3;
   t3 = 1.0/2.0*t3;
   t3 = t3+y0;
   t1 = 1.0/2.0*t1;
   t1 = t1+t3;
   y1 = t1;
   z1 = t2;
   err = abs(y1 - z1);
   return y1;
}
\end{lstlisting}

\subsubsection{\FortranXC{} Implementation}
\lstset{language=fortran}
We want to implement the program of the previous section in \FortranXC{}
from the same computer algebra output
such that all different methods can be used simultaneously in one program.
On top we need to distinguish between the two types `\lstinline!S!' and
`\lstinline!T!', since the latter denotes a three-vector in the case
of the Lorenz oscillator. We introduce a kind
`\lstinline!ki!' for all floating point variables in use:
\begin{lstlisting}[language=fortran]
integer, parameter :: ki = kind(1.0d0)
\end{lstlisting}

The first few lines of the configuration file are modified to accommodate
the differences between \texttt{C} and \FortranXC{}
(see Listing~\ref{lst:rk/fortran.in}).
\begin{lstlisting}[float,language={haggies},frame=tb,%
	caption={[Runge-Kutta Example: \texttt{fortran{.}in}]%
	The modifications made for the file \texttt{rk/fortran{.}in}
   The left out part of the file is identical to \texttt{rk/c{.}in}},%
	captionpos=b,%
	label={lst:rk/fortran.in}]
@language form -> fortran90;
@type T; S;
@coerce
   @int -> S = "%s.0_ki";
   @int/@int -> S = "%s.0_ki/%s.0_ki";
<+$\vdots$+>      <+$\vdots$+>
\end{lstlisting}

One notices that we have not specified the type representation for
`\lstinline!S!' and `\lstinline!T!'. These representations are set
in the template file `\texttt{fortran{.}out}'
(Listing~\ref{lst:rk/fortran.out}) by the \lstinline![% @with %]!
statement that encloses the whole file.
\begin{lstlisting}[float,language={[fortran]haggiest},frame=tb,%
	caption={[Runge-Kutta Example: \texttt{fortran{.}out}]%
	Parts of the template file `\texttt{fortran{.}out}' %
	to produce \FortranXC{} output for the %
	Runge-Kutta example.},%
	captionpos=b,%
	label={lst:rk/fortran.out}]
[% @with env TType="real(ki), dimension(3)"
             SType="real(ki)" %]
   <+$\vdots$+>
function     rk_[%
   @if output.is.file %][%
      output.file match=".*method_(.*).f90"
         format="%s"%][%
   @else %]noname[%
   @end @if%](f, x0, y0, h, err) result(y1)
   <+$\vdots$+>
   [% SType %], intent(in) :: x0, h
   [% TType %], intent(in) :: y0
   [% SType %], intent(out) :: err
   [% TType %] :: y1, z1[%
   @for symbols match="\\$(\\d+)" format="t%s" %]
   [% @select type.name 
      @case T%][% TType %][%
      @case S%][% SType %][%
      @end @select%] :: [% $_ %][%
   @end @for symbols %]
   <+$\vdots$+>
[% @end @with env%]
\end{lstlisting}

The general structure of a \lstinline![% @with %]! statement
is as follows.
\begin{lstlisting}[language={[c]haggiest}]
[% @with <+$name$+> <+$opt_1$[+>=<+$value_1$]+> <+$opt_2$[+>=<+$value_2$]%
       +> <+$\ldots$+>
	   %]body[%
   @end @with %]
\end{lstlisting}
Within its body, symbols are locally defined depending on the
environment which is chosen by $name$ and the options.

\lstset{language=[c]haggiest}
The simplest case is the environment `\lstinline!env!' which takes
its options literally as assignments. Hence we define the
symbols \lstinline![%SType%]! and \lstinline![%TType%]! to denote the
\FortranXC{} representation of the according data type.
Instead of specifying the parameters inside the template
one can load the parameters from a separate file by using the
option `\lstinline!file="properties-file"!' with the `env' environment, where
`\lstinline!properties-file!' must correspond to the name of an existing file.

The second major modification of the template file is in the definition
of the symbols. Instead of the variable \lstinline![% type.repr %]!
a \lstinline![% @select %]! statement is used to choose the
appropriate variable according to the value of \lstinline![% type.name %]!,
which has either the value `\lstinline!T!' or `\lstinline!S!'.

Finally, this template file also demonstrates how identifiers can be
derived from the name of the output file. Since the output might go to
the standard output channel rather than a file, depending on the parameters
with which \haggies{} was called, we need to place an \lstinline![% @if %]!
statement around any block that is using the variable
`\lstinline!output.file!'. Here the if-statement tests if the variable
`\lstinline!output.is.file!' is set to true, which is only the case if
the output is written to a file. The pattern in the option `\lstinline!match!'
selects the relevant part of the file name.
The general syntax of the if-statement is as follows:
\begin{lstlisting}[language={[c]haggiest}]
[% @if <+$name$+> <+$opt_1$[+>=<+$value_1$]+> <+$opt_2$[+>=<+$value_2$]%
       +> <+$\ldots$+> %]branch 1[%
   @elif <+$name_2$+> <+$opt_{21}$[+>=<+$value_{21}$]+> <+%
       $opt_{22}$[+>=<+$value_{22}$]+> <+$\ldots$+> %]branch 2[%
      <+$\vdots$+>
   <+$[$+>@else %]branch <+$n$+>[% <+$]$+>
   @end @if %]
\end{lstlisting}
The if-statement may contain any number of elif-branches and at most one
else-clause.

\begin{figure}[htbp]
\begin{center}
\begin{tabular}{cc}
\includegraphics[width=0.45\textwidth]{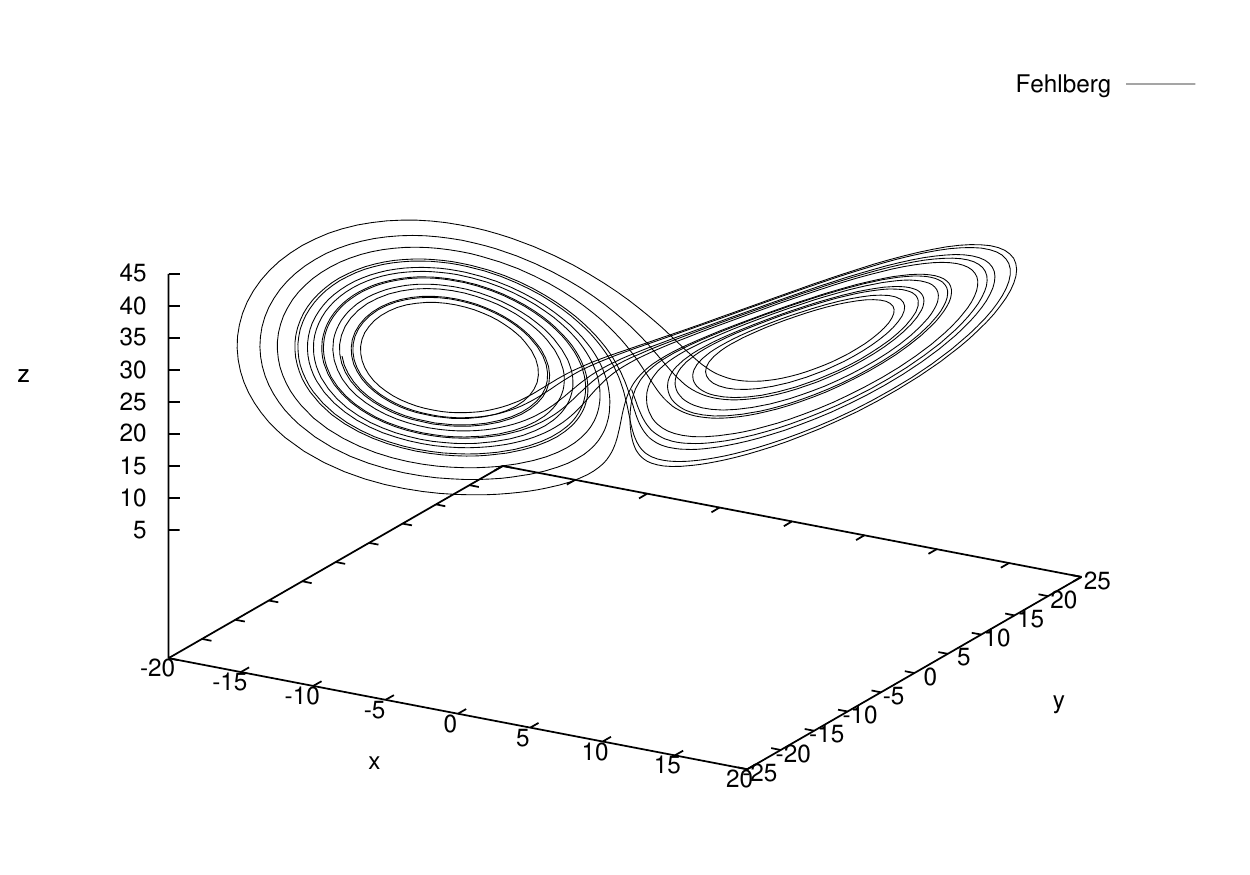}
\includegraphics[width=0.45\textwidth]{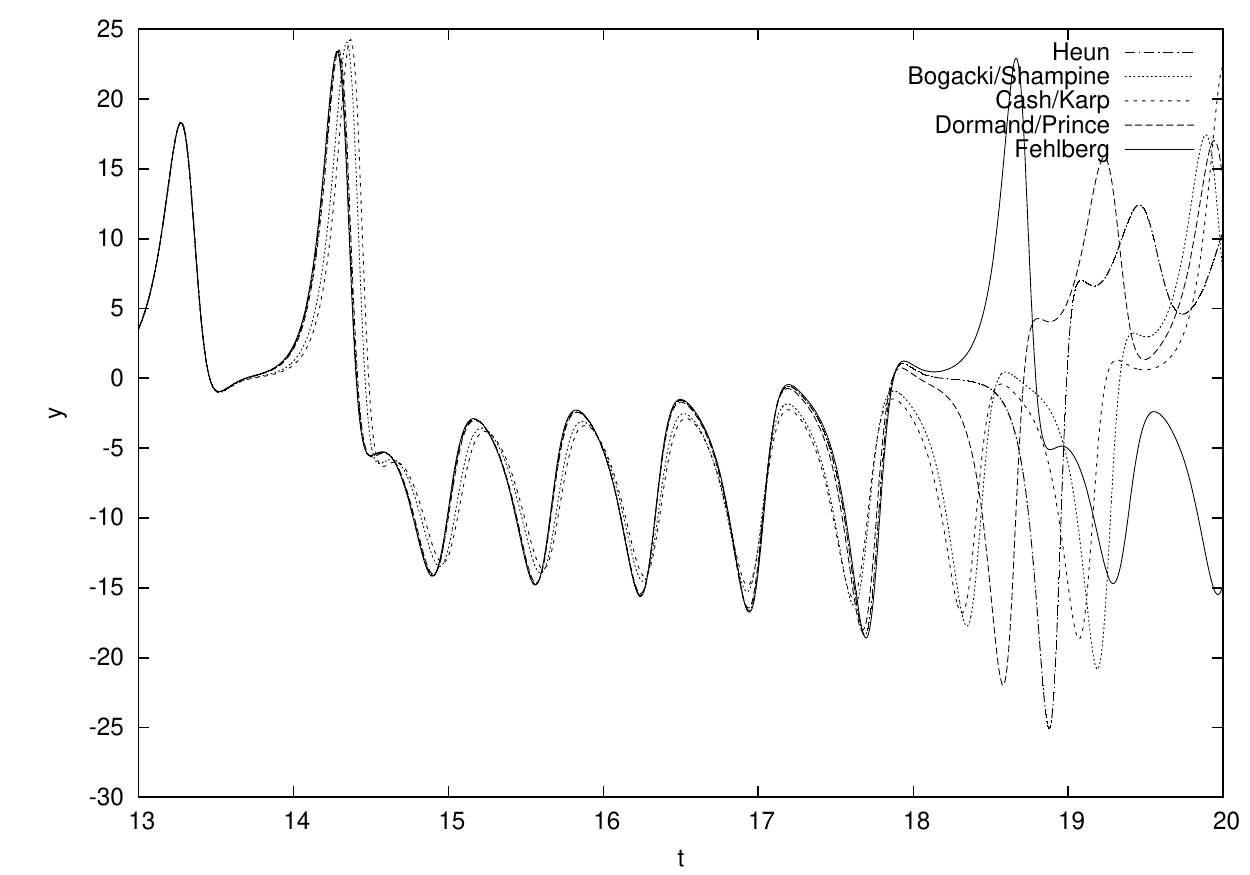}
\end{tabular}
\end{center}
\caption{Results obtained for the Lorenz oscillator as
specified in Equation~\eqref{eq:lorenz-osc} for $0\leq t\leq20$
with $\vec{y}(0)=(0,1,20)$ and a precision of $10^{-6}$.
The graphics on the left shows the solution integrated with the
Runge-Kutta-Fehlberg method, which is fourth order.
The second graph compares the evolution of the
second coordinate for different methods.
Both plots are generated for a goal precision of $10^{-6}$.
At $t>18$ the integration becomes unreliable. For $10^{-7}$ (not shown)
all curves would follow the Fehlberg method, which does not
change substantially by increasing the precision beyond~$10^{-6}$.}\label{fig:lorenz}
\end{figure}

The directory contains a \texttt{make} file which generates the file
\texttt{lorenz.exe} along with the \FortranXC{} files for all implemented
Runge-Kutta methods. The program \texttt{lorenz.f90} implements an adaptive
integrator that uses the different Runge-Kutta methods for the time step.
In Figure~\ref{fig:lorenz} the different methods are compared for a
common precision goal.

\subsection{Gaussian Quadrature}\label{ssec:examples:op}
\begin{tabular}{lp{0.6\textwidth}}
\bf Synopsis: & Obtaining weights and roots for Gaussian quadrature
    with non-standard integration kernels \\
\bf Objective: & output format for user-defined types in languages
    without operator overloading \\
\bf Requirements: & \FORM{}, Java compiler \\
\bf Directory: & \texttt{demo/op}
\end{tabular}

\subsubsection{Overview}
Gaussian quadrature describes the systematic approximation of an
integral by a finite sum. In this example we restrict the formulae to the
integration interval between zero and one as in
\begin{equation}\label{eq:gauss:def01}
\int_0^1\mathrm{d}x\;\omega(x)f(x)\approx\sum_{i=1}^N\omega_if(x_i)\text{,}
\end{equation}
where $\omega(x)$ denotes a distribution and $f(x)$ a function which is
regular in the integration region. The extension to distributions is
motivated by the regularisation of infrared singularities in gauge
theories. The singularity structure of scattering amplitudes at
higher orders leads one to integrals of the form
\begin{align}
I_1&=\int_0^1\mathrm{d}x\;\left(-\frac{\log(1-x)}{1-x}\right)_+f(x)%
\text{,}\\
I_2&=\int_0^1\mathrm{d}x\;\left(\frac1{1-x}\right)_+f(x)
\end{align}
and similar integrals, where the plus-distribution $()_+$ denotes the
prescription
\begin{equation}
\int_0^1\mathrm{d}x\;\left(g(x)\right)_+f(x)\equiv
\int_0^1\mathrm{d}x\;g(x)\left(f(x)-f(1)\right)\text{.}
\end{equation}
Since the function $f(x)$ itself is related to a cross-section and
needs to be evaluated through Monte-Carlo integration a typical strategy
is the evaluation of the $x$-integral through Monte-Carlo techniques which
requires the approximation of the distributions $\omega(x)$ by functions
that are suitable for Monte-Carlo integration. These technical difficulties
could be avoided if the $x$-integration is carried out by a quadrature rule
that is adapted to the integration kernel. A drawback, of course, is that
the evaluation of the Monte-Carlo integrals in $f(x)$ has to be repeated
a fixed number of times at positions $x_i$;
on the other hand any problems connected to the
approximation of the distributions can be avoided in the case of a
Gauss quadrature for the $x$-integral and one expects a better
convergence of the Monte-Carlo integrals inside $f(x_i)$.

The determination of the weights $\omega_i$ and the roots $x_i$ in
Equation~\eqref{eq:gauss:def01} requires the introduction of
polynomials $p_m(x)$ orthogonal with respect to the inner product
\begin{equation}
\langle p_m,p_n\rangle=\int_0^1\mathrm{d}x\;\omega(x)p_m(x)p_n(x)=\delta_{mn}%
\text{.}
\end{equation}
The form of these polynomials can be derived using the determinant
formula below.\footnote{The use of $(1-x)$ instead of $x$ is motivated
by the form of $\omega(x)$.}
\begin{equation}\label{eq:gauss:determinant}
p_n(x)\propto\left\vert
\begin{array}{ccccc}
\sigma_{0,0}&\sigma_{0,1}&\cdots&\sigma_{0,n-1}&1\\
\sigma_{1,0}&\sigma_{1,1}&\cdots&\sigma_{1,n-1}&(1-x)\\
\sigma_{2,0}&\sigma_{2,1}&\cdots&\sigma_{2,n-1}&(1-x)^2\\
\vdots&\vdots&\ddots&\vdots&\vdots\\
\sigma_{n-1,0}&\sigma_{n-1,1}&\cdots&\sigma_{n-1,n-1}&(1-x)^{n-1}\\
\sigma_{n,0}&\sigma_{n,1}&\cdots&\sigma_{n,n-1}&(1-x)^{n}
\end{array}
\right\vert
\end{equation}
Here we used the symbols
\begin{equation}
\sigma_{ij}=\int_0^1\!\!\mathrm{d}x\omega_p(x)(1-x)^{i+j},\quad
\forall i,j\geq0
\end{equation}
which for the above kernels are
\begin{align}
\sigma_{0,0}^{(1)}&=0\quad\text{and}\quad
\sigma_{i,j}^{(1)}=\frac1{(i+j)^2},\quad i+j>0\\
\sigma_{0,0}^{(2)}&=0\quad\text{and}\quad
\sigma_{i,j}^{(2)}=\frac1{i+j},\quad i+j>0
\end{align}

The roots $x_i$ in Equation~\eqref{eq:gauss:def01} 
are the roots of the polynomial $p_N$ and the weights $\omega_i$
are defined as
\begin{equation}
\omega_i=\int_0^1\!\!\mathrm{d}x\,\left(\omega(x)\right)_+\prod_{j\neq i}%
\frac{x-x_j}{x_i-x_j}
\text{.}
\end{equation}

\subsubsection{Implementation}
We use the equations from the previous section to generate the expressions
for $p_N(x)$ for different values of $N$ using a
\Form{}~\cite{Vermaseren:2000nd,Vermaseren:2006ag} program that
evaluates the determinant in Equation~\eqref{eq:gauss:determinant}.
The program \haggies{} is used to produce a \texttt{Java} program which
evaluates $\omega_i$ and $x_i$ to arbitrary precision using the class
\texttt{BigDecimal} for the numerical operations.

At this point, one of the shortcomings of \texttt{Java} is the lack of
operator overloading. Rather than simply writing \texttt{x*y} one has
to use the form \texttt{x.multiply(y)} if \texttt{x} and \texttt{y}
are declared as objects of the class \texttt{BigDecimal}, which in
\texttt{Java} implements arbitrary precision numbers.

In \haggies{} this problem is circumvented by using patterns to specify
the form of arithmetic operations for each type. The configuration file
of the integration example is shown in Listing~\ref{lst:op.in}.

\begin{lstlisting}[float,language=haggies,frame=tb,%
	caption={[Configuration file \texttt{op{.}in}]%
   Configuration file \texttt{op{.}in} used in the Gaussian %
   quadrature example. The symbol \texttt{[1-x]} as used in \FORM{}%
   needs to be escaped properly.},label=lst:op.in]
@language form -> c;
@type F = "BigDecimal", "%s.add(%s, mc)",
          "%s.subtract(%s, mc)", "%s.negate(mc)";
@coerce
	@int -> F = "number(\"%s\", mc)";
	@int/@int -> F = "number(\"%s\", \"%s\", mc)";
@define
	\[1\-x\] : F = "xbar";
	N, P : F;
@operator
	F * F -> F = "%s.multiply(%s, mc)",
                "%s.divide(%s, mc)";
@polynomial \[1\-x\];
\end{lstlisting}

\lstset{language=haggies}
The most obvious change to the previous examples are the additional arguments
to the \lstinline!@type! and \lstinline!@operator! statements.
For the type-statement the arguments from left to right have the meaning
\begin{enumerate}
\item type name in the target language; this value can be accessed
      through the field \lstset{language=[c]haggiest}
      \lstinline![% type.repr %]!
      in the template file
\item pattern for binary `$+$' operator
\item pattern for binary `$-$' operator
\item pattern for unary `$-$' operator
\end{enumerate}
\lstset{language=haggies}
The patterns use the same syntax as the \texttt{format} method in
\texttt{Java}. This means that in order to change the order of the
operands one could use the form \lstinline!"%2$s ... %1$s"! to select
the second operand before the first one.
Similarly, the arguments to the \lstinline!@operator! statements
are in this order
\begin{enumerate}
\item pattern for multiplication
\item pattern for division
\item pattern for inverse division
\end{enumerate}
The third argument becomes relevant if two different types are involved.
Consider, for example, the expression 
$t/s$ with $s:\mathrm{S}$ and $t:\mathrm{T}$.
This would match the operation \lstinline!S*T->R! although the operands
are in reverse order. In general, we cannot assume that both $s/t$ and $t/s$
can be represented by the same pattern in the target language.
Hence, \haggies{} provides two different patterns. However, one must
take care to use the right order of the arguments in the patterns.
The definition
\begin{lstlisting}[language=haggies]
@operator S*T -> R = "mul_S_T(%s, %s)",
      "div_S_T(%s, %s)", "div_T_S(%2$s, %1$s)";
\end{lstlisting}
translates $s/t$ into \texttt{div\_S\_T(s, t)} and
$t/s$ into \texttt{div\_T\_S(t, s)}. Omission of the third
argument would lead to \texttt{div\_S\_T(t, s)} in the case of $t/s$.

Another important issue about using the \texttt{BigDecimal} class is
that numeric literals cannot be constructed at compile time. The
number~$4.0/7.0$, for example is constructed by the expression
\begin{lstlisting}[language=Java]
(new BigDecimal("4.0")).divide(new BigDecimal("7.0"))
\end{lstlisting}
To avoid unnecessary constructor calls, \haggies{} provides the command
line option\footnote{The long form is \texttt{--cse-on-numbers}.} \texttt{-n}
which causes the \ac{CSE} to treat numbers as symbols. Therefore, in
the expression $2/3x+2/3y$ with the option \texttt{-n} activated the number
$2/3$ is replaced by a symbol and reused in both terms.

As an example result Table~\ref{tbl:op:numbers} shows the roots and weights
for $I_1$ and $I_2$ for $n=10$.

\begin{table}[hbt]
\begin{center}
\tiny
\begin{tabular}{r|rr|rr}
& \multicolumn{2}{c|}{$I_1$} & \multicolumn{2}{c}{$I_2$} \\
\multicolumn{1}{c|}{$i$} &
\multicolumn{1}{c}{$x_i$} & \multicolumn{1}{c|}{$\omega_i$} & 
\multicolumn{1}{c}{$x_i$} & \multicolumn{1}{c}{$\omega_i$} \\
\hline
1 & 0.03426906910338113 & 0.004059350690970753 &
0.01441240964045317 & 0.03968369730743715\\
2 & 0.1117456505243587 & 0.003451095443491922 &
0.07438738971914130 & 0.08034566242558715 \\
3 & 0.2249228996671723 & 0.06790187522156536 &
0.1761166561583394 & 0.1627776617949430 \\
4 & 0.3632281380485790 & 0.04686069099597275 &
0.3096675799277467 & 0.1797579743081136 \\
5 & 0.5137950092446798 & 0.3399527393234391 &
0.4619704010795133 & 0.3414395624615283 \\
6 & 0.6626937961738814 & 0.2391750197385724 &
0.6181172346954572 & 0.3199520268272458 \\
7 & 0.7962756988248819 & 1.400458498733316 &
0.7628230151850353 & 0.7002392453690284 \\
8 & 0.9025071571982903 & 1.097197140776207 &
0.8819210212099773 & 0.6107993057794040 \\
9 & 0.9721994578865998 & 9.795036227495281 &
0.9637421871167904 & 2.322941371637727 \\
10 & 1.000000000000000 & -12.99409263842344 &
1.000000000000000 & -4.757936507922282
\end{tabular}
\end{center}
\caption{The roots $x_i$ and the weights $\omega_i$ for the integration
kernels $\omega(x)$ of the integrals $I_1$ and $I_2$.}
\label{tbl:op:numbers}
\end{table}

\subsection{Colour Algebra}\label{ssec:examples:color}
\begin{tabular}{lp{0.6\textwidth}}
\bf Synopsis: & Calculating the colour correlation matrix for a
    \acs{QCD} matrix element in a certain basis \\
\bf Objective: & Initialisation of a matrix\\
\bf Requirements: & \FORM{} \\
\bf Directory: & \texttt{demo/color}
\end{tabular}
\subsubsection{Overview}
The previous two examples already introduced all important language
features present in \haggies{}. This example is taken from the \GOLEM{}
project; it generates the colour correlation matrix for the six-gluon
amplitude in \ac{QCD}. If one assigns the adjoint $\mathsf{SU}(3)$ indices
$A_1,\ldots,A_6$ to the gluons, a colour basis can be defined by the
traces of the form
\begin{subequations}
\begin{align}
[120]\quad&\mathrm{tr}\{t^{A_{\sigma_1}}\cdots t^{A_{\sigma_6}}\},\\
[90]\quad&\mathrm{tr}\{t^{A_{\sigma_1}}t^{A_{\sigma_2}}\}
\mathrm{tr}\{t^{A_{\sigma_3}}\cdots t^{A_{\sigma_6}}\},\\
[40]\quad&\mathrm{tr}\{t^{A_{\sigma_1}}t^{A_{\sigma_2}}t^{A_{\sigma_3}}\}
\mathrm{tr}\{t^{A_{\sigma_4}}t^{A_{\sigma_5}}t^{A_{\sigma_6}}\},\\
[15]\quad&\mathrm{tr}\{t^{A_{\sigma_1}}t^{A_{\sigma_2}}\}
\mathrm{tr}\{t^{A_{\sigma_3}}t^{A_{\sigma_4}}\}
\mathrm{tr}\{t^{A_{\sigma_5}}t^{A_{\sigma_6}}\},
\end{align}
\end{subequations}
where $\sigma$ in each case
are the distinguishable permutations of the indices $\{1,\ldots,6\}$.
The numbers in square brackets denote how many of these permutations
exist. Adding up these numbers yields the dimension of the colour
space for this process, which is 265. If we denote these basis elements
by $\kea{i}$, $i=1,\ldots 265$ the contraction of the adjoint indices
between a basis element and a conjugate basis element
defines an inner product $\Spaa{ij}$.
The corresponding Gram matrix is called the colour correlation matrix
and plays a role in squaring \ac{QCD} amplitudes.

\subsubsection{Implementation}
This example contains a \FORM{} program which constructs all $\Spaa{ij}$
as functions in $N_C$, the number of colours, which is $3$ in \ac{QCD}.
These entries are stored in variables \texttt{CC\_$i$\_$j$} in the file
\texttt{color.txt}. The constant \texttt{TR} is the normalisation
of the generators $\mathrm{tr}\{t^At^B\}=T_R\delta^{AB}$. \haggies{}
is used to transform this file into a \FortranXC{} program to initialise
an array $\mathtt{CC}(i,j)=\Spaa{ij}$. Since many of the entries of
$\Spaa{ij}$ are zero it would be very inefficient to generate an
instruction of the form
for each zero entry. A more efficient solution is to initialise
the whole array to zero and to modify only the non-zero entries.
This is achieved by the following fragment in the file \texttt{color.out}:
\begin{lstlisting}[language={[c]haggiest}]
[% @for instructions %][%
   @select $_ match="(.).*" format="%s"
   @case $ %]
[% $_ match="\\$(\\d+)" format="t%s" %] = [%
     expression match="\\$(\\d+)" format="t%s" %][%
   @case C %][% @if is_zero %][% @else %]
[% $_ match="CC_(\\d+)_(\\d+)"
       format="CC(%s, %s)" %] = [%
     expression match="\\$(\\d+)" format="t%s" %]
[%     $_ match="CC_(\\d+)_(\\d+)"
       format="CC(%2$s, %1$s)" %] = [%
     expression match="\\$(\\d+)" format="t%s" %][%
     @end @if is_zero %][%
   @else %]
[% $_ %] = [%
     expression match="\\$(\\d+)" format="t%s" %][%
   @end @select %][%
   @end @for instructions %]
\end{lstlisting}
There are two points worth mentioning in this template file:
\begin{itemize}
\item The symbol `\lstinline!is_zero!' inside a `\lstinline!instructions!'
   loop evaluates to true if the right-hand side of the assignment is
   zero. Hence the corresponding if-statement in the example above
   only produces code for those assignments with non-zero right-hand side.
\item The variables \texttt{CC\_$i$\_$j$} are transformed by the
   match-format pair
\begin{lstlisting}[language={[fortran]haggiest}]
[%$_ match="CC_(\\d+)_(\\d+)" format="CC(%s, %s)"%]
\end{lstlisting}
   which stores the current values of $i$ and $j$ in the two groups\footnote{%
   The pattern \lstinline!\\d! matches a decimal digit}
   `\lstinline!(\\d+)!' and uses them to fill the two fields
   `\lstinline!%s!'.
\end{itemize}

\subsection{Amplitude Calculations}\label{ssec:examples:cut}
\begin{tabular}{lp{0.6\textwidth}}
\bf Synopsis: & Preparing a Feynman-diagrammatic expression
      as input for the OPP method\\
\bf Objective: & Using brackets and exploiting possibilities
                 for caching \\
\bf Requirements: & \FORM{}, \FortranXC{} Compiler \\
\bf Directory: & \texttt{demo/cut}
\end{tabular}
\subsubsection{Overview}
In this example we consider one of the main original motivations behind
\haggies{}, the evaluation of one-loop scattering amplitudes. We outline
a method different from the one implemented in \GOLEM{}, which suits
better as a short example.

A one-loop amplitude\footnote{For the ease of the argument we consider
leg-ordered colour-subamplitudes.}
can be represented in a function basis consisting of
scalar Feynman integrals
\begin{equation}\label{eq:basisfunctions}
\mathcal{M}_{\text{virt}}=\sum_{N=1}^{4}
\sum_{\alpha}c_{N,\alpha} I_N^n(S^\alpha)+\mathcal{R}\text{,}
\end{equation}
where the functions $I_N^n(S)$ are defined in
Equation~\eqref{eq:define-tensint} and the label $\alpha$ runs over
all possible deletions of propagators that lead to the according $N$-point
topologies. The last term $\mathcal{R}$ is the so-called rational term
which is free of transcendental functions.

As mentioned in the introduction, in the \GOLEM{} approach one uses
recurrence relations to reduce the amplitude to a basis similar to the
one in Equation~\eqref{eq:basisfunctions}~\cite{Binoth:2005ff}.
An alternative approach is to use the knowledge about the structure of the
amplitude to determine the coefficients $c_{N,\alpha}$, which the
OPP method achieves by evaluating the integrand under the
$\int\!\!\mathrm{d}^nk$ integral for specific values
of~$k$~\cite{Ossola:2006us}. One of the drawbacks of this method is
the fact that $\mathcal{R}$ can only be reconstructed partially and
another method is required for its full determination. Therefore ---
and for the sake of simplicity --- the example program only deals with
the cut constructable parts of the amplitude and leaves out the discussion
of $\mathcal{R}$.

\subsubsection{Implementation}
In order not to overload the example by physics details, the input
expression has been chosen as the numerator of a box diagram from the
reaction $u\bar{u}\rightarrow t\bar{t}$, contracted with the tree-level
diagram. This contraction ensures that all spinor lines can be expressed
as traces, which can be expanded efficiently in \FORM{}. The resulting
expression is written in terms of Mandelstam variables $s_i=k_i^2$ and%
\footnote{All external momenta $k_i$ are incoming.}
$s_{ij}=(k_i+k_j)^2$ and in terms of dot-products involving the integration
momentum~$q$, such as $q^2$ and $q\cdot k_i$.

The authors of~\cite{Ossola:2006us} provide a \FortranXC{} implementation
of their algorithm (\texttt{CutTools}) which is described
in~\cite{Ossola:2007ax}. As an input this program requires a function
$\mathtt{num}(k, \tilde{k}^2)$ which evaluates the numerator of the
amplitude for a specific value of the integration momentum,
a complex four-vector $k$, and the value of the $(n-4)$ dimensional part
$\tilde{k}^2$.

The \FORM{} program \texttt{numerator.frm} generates an expression for the
numerator as described; then we use \haggies{} to write the \FortranXC{}
routine \texttt{num} in a form where the $k$ dependence is separated from
the dependence on $s_i$ and $s_{ij}$, such that large parts of the expression
are evaluated only once per phase space point and not for each value of $k$.
Since the set of Mandelstam variables depends on the number of external
particles the configuration file \texttt{numerator.in} and the according
definitions in \FortranXC{} are written by the \FORM{} program, allowing
for different processes to be calculated by just small modifications.

In the input file we focus on the two lines below and their implications
for the rest of the program:
\begin{lstlisting}[language=haggies]
@brackets [1] q, q2;
@brackets [2] 3000;
\end{lstlisting}
The first line advises \haggies{} to factor out all occurrences of
the vector~\texttt{q}, which in the implementation represents the
integration momentum~$k$,
and of the symbol $\mathtt{q2}=k^2$. This statement will also factor
out all dot-products involving~\texttt{q}.

The second line introduces a second level of brackets wherever a subexpression
contains more than 3000 terms, which has been chosen arbitrarily and should
be changed to much smaller values if the user wants to explore its effect for
this simple example.
It should be noted that, in order to avoid unnecessary function calls,
\haggies{} does not introduce this
extra level of brackets if only one bracket would be generated.

In several places of the template file \texttt{numerator.out}
we find a pair of nested loops as the one shown below
\begin{lstlisting}[language={[fortran]{haggiest}}]
[% @do brackets prefix="outer." bracket="subex%03d" %]
[%    @do brackets prefix="inner."
             bracket="subex%2$03d_%1$03d" %]
   ! inner block
[%    @end @do %]
   ! outer block
[% @end @do %]
\end{lstlisting}
\lstset{language={[fortran]{haggiest}}}
The two different prefixes have been chosen since the variables of
the outer block should also be visible inside the inner block.
The option `bracket' determines how the
how the name of the bracket is displayed. For each (nested) bracket
\haggies{} keeps a path of numbers which uniquely determines the
current bracket. The outermost loop corresponds to the
command \lstset{language=haggies}\lstinline!@brackets[1]!
in the configuration file, whereas the nested loop corresponds to
the command \lstinline!@brackets[2]!.
\lstset{language={[fortran]{haggiest}}}
The values of `\lstinline!outer.$_!' have the values \texttt{subex001},
\texttt{subex002} etc; the values of `\lstinline!inner.$_!' are
\texttt{subex001\_001}, \texttt{subex001\_002} etc where
`\lstinline!outer.$_!' is \texttt{subex001} and \texttt{subex002\_001} etc
for the second entry of the outer loop and so on. This means, the fields
are filled left to right from the innermost to outermost bracket level.
This behaviour makes it easier to access the innermost counter
in languages where local functions can be used for nested brackets.

\lstset{language=[c]{haggiest}}
Inside each brackets-loop the variable `\lstinline!type.repr!'
denotes the type of the current bracket and the user has also
access to the iterators `\lstinline!symbols!' and `\lstinline!instructions!'
which iterate over the local symbols (resp. instructions) of the
subprogram which calculates the current bracket. The result of the
bracket is stored in the symbol `\lstinline!&result!'.

In order to propagate the results of the bracket from the inner
to the outer loops the function \lstinline![% expression %]!
has the argument `\lstinline|bracket|', which must be set
appropriately if the current expression can refer to a bracket.
The declaration part of the function \texttt{num} looks as follows
in the template file:
\begin{lstlisting}[language={[fortran]{haggiest}}]
[% @for symbols %][% 
     type.repr %] :: [% $_
         match="\\$(.*)" format="r%s" %]
[% @end @for symbols %][%
   @for brackets prefix="outer." bracket="b%d" %]
[% outer.type.repr %], save :: [% outer.$_ %] = &
&    (0.d0, 0.d0)[%
   @end @for brackets %]
\end{lstlisting}
The first loop introduces local variables for all temporary
variables. The second loop declares local, static variables \texttt{b1},
\texttt{b2}, \dots{} for the brackets from the first bracket level.

In order to recompute the values of the variables \texttt{b}$i$ only
when needed the flag \texttt{dirty\_cache} is introduced, which is
set whenever the kinematics of the process is updated. The following
lines set up a conditional block for the recomputation of these symbols.
\begin{lstlisting}[language={[fortran]{haggiest}}]
[% @for brackets prefix="outer."
                 bracket="subex%03d" %][%
   @if outer.is_first %]if (dirty_cache) then[%
   @end @if %]
   b[% outer.index %] = [% outer.$_ %]()[%
   @if outer.is_last %]
   dirty_cache = .false.
end if[%
   @end @if %][%
@end @for brackets %]
\end{lstlisting}
We use the `\lstinline!is_first!' and `\lstinline!is_last!' functions
to put the if-block in the right place\footnote{This also ensures that no
empty if-statement is generated, if no brackets are present}. Inside
the loop we need to access the symbols \texttt{b1}, \texttt{b2}, etc
at the same time as the symbols \texttt{subex001}, \texttt{subex002}~etc.
Since there is only one `\lstinline!bracket!' option to specify a pattern
we construct the left-hand sides explicitly by accessing the function
`\lstinline!outer.index!'.

The last part of the subprogram is the loop over the instructions.
The main difference to any of the previous examples is the additional
option `\lstinline!bracket!' in the function `\lstinline!expression!'
at the right-hand side of the assignments. This option is used to
format the brackets in the generated expressions.
It should be noted that here the pattern which is used
is `\lstinline!"b%d"! which refers to the cached symbols rather than
the functions to calculate the brackets.
\begin{lstlisting}[language={[fortran]{haggiest}}]
q2 = dotproduct(q, q)[%
   @for instructions %]
[% @select $_ match="(.).*" format="%s"
   @case $ %][% $_ match="\\$(.*)" format="r%s" %][%
   @else %][% $_ %][%
   @end @select %] = [% expression
       match="\\$(.*)" format="r%s"
       bracket="b%d" %][% 
   @end @for instructions %]
\end{lstlisting}

The subroutines for the brackets are structured in a similar way with
one main difference: the caching is applied only at the
outermost level since in this problem there are only two sets of symbols
with respect to their update-frequency, the integration momentum $k$
which is different for each call and the Mandelstam variables which
only change for each new external kinematics.
However, for other problems it might well make sense to introduce multiple
levels of caching, for example in multidimensional parameter scans. 

\subsubsection{Parallelisation}
In the file \texttt{parallel.out} we introduced the necessary
OpenMP~\cite{openMP} directives which are necessary to parallelise
the computation of the brackets. The only lines which are different
from the above code are inside the initialisation block for the
variables \texttt{b}$i$:
\begin{lstlisting}[language={[fortran]{haggiest}}]
[% @for brackets prefix="outer."
                 bracket="subex%03d" %][%
   @if outer.is_first %]if (dirty_cache) then
   !$OMP PARALLEL SECTIONS[%
   @end @if %]
   !$OMP SECTION
   b[% outer.index %] = [% outer.$_ %]()[%
   @if outer.is_last %]
   !$OMP END PARALLEL SECTIONS
   dirty_cache = .false.
end if[%
   @end @if %][%
@end @for brackets %]
\end{lstlisting}
The inserted directives create a parallel section for each of the
function calls inside the if-statement. The actual performance gain,
of course, depends very much on the problem. For simple cases almost
always the overhead of creating the threads is larger than the speedup
from the parallel calculation. Depending on the number and the sizes
of the brackets one might also consider other than the outermost level
to introduce parallelism.

\subsection{Interval Constraint Solver}\label{ssec:examples:cs}
\begin{tabular}{lp{0.6\textwidth}}
\bf Synopsis: & Solving the Broyden-Banded function\\
\bf Objective: & Generating output for a scripting language,
      working with interval arithmetic\\
\bf Requirements: & \FORM{}, Python \\
\bf Directory: & \texttt{demo/cs}
\end{tabular}
\subsubsection{Overview}
This last example implements an interval constraint solver. The basic
idea here is to determine the zeroes of a set of functions
\begin{equation}
\vec{x}\,\text{such that}\,f_i(\vec{x}) = 0,\quad\forall i=1,\ldots,n
\end{equation}
by a so-called \emph{branch and prune} algorithm~\cite{Hentenryck}.

The program uses interval arithmetic to map a box $B\subset \mathds{R}^n$
onto a box $f(B)$; if $\vec{0}\in f(B)$ the box $B$ is divided into halves
along one of the dimensions, $B_1\cup B_2=B$, and the algorithm is
applied to both of the daughter boxes. If $\vec{0}\not\in B$ the box
is discarded. The algorithm stops if each of the boxes has a small enough
volume and is considered a solution,
which must be checked by another method, since from $\vec{0}\in f(B)$
one cannot conclude $\exists \vec{x}\in B: f(\vec{x})=\vec{0}$ if $f(B)$
is calculated by the rules of interval arithmetic.

We implement the method for the Broyden banded function, a typical benchmark
for optimisation software~\cite{More:1981}. This set of functions is defined
as
\begin{equation}
f_i(x_1,\ldots, x_n)=x_i(2+5x_i^2) + 1 - \sum_{j\in J_i} x_j(x_j+1)
\end{equation}
where $J_i=\{j:j\neq i,\max(1,i-5)\leq j\leq\min(n,i+1)\}$.

\subsubsection{Implementation}
A simple \FORM{} program generates the set of equation for a given $n$.
The output is read by \haggies{} and transformed into a Python program.
The interval arithmetic is implemented by the package \texttt{mpmath}
by Fredrick Johansson\footnote{\url{http://mpmath.googlecode.com}}
As in the first example, we specify the option \texttt{-n} to \haggies{}
at the command line to avoid multiple generation of interval objects of
constants. In the template file the assignment to the variables $f_i$
is replaced~by
\begin{lstlisting}[language={[python]haggiest}]
if 0 not in [% expression ... %]:
   return False
\end{lstlisting}
This block returns from the function and yields \texttt{False} if any
of the expressions can be proved not to contain zero.

\clearpage
\section{Conclusion}\label{sec:conclusion}

In this article we have reported on \haggies{},
an open source program
for the generation of optimised code for the efficient numerical
evaluation of mathematical expressions.
We have put an emphasis on the generality of the code generator, restricting
the set of possible output formats as little as possible. Part of this
effort is the introduction of a type inference system which allows the
user to combine different data types to represent various algebraical objects.
The example programs demonstrate that the presence of such a type
inference system provides the capability of handling
situations which are difficult to cope with
using traditional code generators. On the one hand, it allows to declare
temporary variables with the proper type in statically typed languages
without the use of implicit typing\footnote{The keyword \texttt{implicit}
in \texttt{Fortran} is a well known source for undetected programming errors.}.
On the other hand, the type system makes it easy to specify rules
for translating the algebraic operators into function and method calls,
which is required when working
with non-standard types in languages without operator overloading, such
as \texttt{Java}.

The program \haggies{} transforms the input expressions by a
multivariate Horner scheme reducing the number of multiplications
and performs a \acl{CSE} on the resulting expression. The careful
choice of algorithms in the code generator guarantees that the
time for the code transformation scales linearly with the size of
the input expression. This enables us to transform and compile
relatively large expressions: we currently use \haggies{} successfully
inside \GOLEM{}, a package for generating code for one-loop corrections
to scattering processes in quantum field theories. In this application
\haggies{} deals with expressions of up to several 10MB consisting of
up to $\mathcal{O}(10^5)$ terms.

We have implemented a notion of bracketing at an arbitrary number of
levels to factor out certain symbols
treating each bracket in a separate subprogram. We find this feature
useful in three main applications: first of all these brackets allow to
split an expression into smaller, independent subprograms if the resulting
program is too large for compilation. A second advantage is the possibility
of factoring out symbols that change more frequently than others. The
(relatively) constant subexpressions can be stored in static variables
and are only recomputed if needed. A third aspect is parallelisation
as it is very easy to parallelise the subprogram invocations for the
different subexpressions.

The features such as the type inference system
and the bracketing are mandatory for the use of \haggies{} inside
\GOLEM{} and beneficial for many other applications.
Through the separation of the template files from the computer algebra
one can produce different programs from the same source without modifying
the \ac{CAS} code, which also facilitates the integration of \haggies{}
with existing projects.

\section*{Acknowledgements}
The author thanks the organisers
of the Les Houches workshop on Physics at TeV Colliders, where part of
this work has been completed.
The author wants to acknowledge Gudrun Heinrich for careful testing of the
program and for stimulating discussion.

\appendix
\section{Input Description}\label{ssec:reference:in}
\lstset{language=haggies}

\subsection{The \texorpdfstring{\lstinline!@language!}{@language} Statement}%
\label{r:language}
The configuration file must contain exactly one
\lstinline!@language! statement. The exact syntax is
\begin{lstlisting}
@language input_language -> output_language;
\end{lstlisting}
Currently, possible choices for `\lstinline!input_language!' are
`\lstinline!mathematica!' and `\lstinline!form!'. The latter option
is suitable for most computer algebra formats. The Mathematica format
implements a subset of the Mathematica language and allows for the most
commonly used operators. In both cases the file of input expressions
consists of a list of assignments with the equals sign (`\lstinline!=!')
as assignment operator and a semicolon terminating each expression.

Possible values for `\lstinline!output_language!' are
\begin{description}
\item[\texttt{fortran90}] produces output for \FortranXC{} and later
versions of \texttt{Fortran}.
\item[\texttt{c}] is recommended for \texttt{C}-like languages
like \texttt{C/C++}, \texttt{Java}. It is also suitable for many
scripting languages that do not require continuation characters
for continuation lines.
\item[\texttt{python}] can be used for scripting languages which use
the backslash as continuation character. The exponentiation operator
is set to `\texttt{**}' but can be changed in the configuration file.
\item[\texttt{lisp}] produces expressions in prefix notation
for \texttt{Lisp}.
\item[\texttt{mathematica}] produces output suitable for \texttt{Mathematica}.
\item[\texttt{maple}] is recommended for \texttt{Maple}.
\item[\texttt{ada}] is recommended for \texttt{Ada}.
\end{description}

\subsection{The \texorpdfstring{\lstinline!@type!}{@type} Statement}%
\label{r:type}
The definition of a new type is introduced by the
\lstinline!@type! statement. In its simplest form
it consists only of the keyword and the type name.
\begin{lstlisting}
@type T;
\end{lstlisting}
Each type that one wants to use needs to be declared exactly once
in a configuration file.
Optional arguments can be added after an equals-sign to specify
the representation of the type in the target language and the representation
of the operations `$a+b$', `$a-b$' and `$-a$'. Therefore, the following
declarations are valid.
\begin{lstlisting}
@type T1 = "float";
@type T2 = "double[]", "add_vectors(%s,%s)",
"subtract_vec(%s,%s)", "negate_vec(%s)";
\end{lstlisting}
The rightmost arguments can be dropped; however, there are hardly any
cases where some but not all of the operations need to be redefined.
If no right-hand side is specified the type representation defaults to
the type name on the left-hand side.

\haggies{} has two built-in types, \lstinline!@int! and
\lstinline!@int/@int! which are the types of integer and rational
literals respectively.

\subsection{The \texorpdfstring{\lstinline!@define!}{@define} Statement}%
\label{r:define}
All symbols that appear in an expression need to be defined
with a defined type. An optional which determines the representation
in the target language can be added with an equals sign. On the left-hand
side of the definition, a list of patterns is allowed.
\begin{lstlisting}
@define
   x, y, z: T1;
   arr: I -> T2 = "a[%2$s]";
   f, g, h: T1, T1 -> T1 = "%s(%s, %s, dummy)";
\end{lstlisting}
The arrow (\lstinline!->! = \texttt{->}) denotes a function type.
Currently, no overloaded or generic functions can be specified
and the number of arguments needs to be fixed.
When defining functions the first wildcard in the pattern contains
the function name and the arguments start with the second wildcard.
As the wildcard syntax follows \texttt{Java}'s syntax for the class
\texttt{java.util.Formatter}, one can explicitly access the different
arguments by their number (starting from the index 2): in order to
transform $f(a, b, c)$ into $a.f(c, b)$ one can use the pattern
\lstinline!"%2$s.%1$s(%4$s, %3$s)"!.

\subsection{The \texorpdfstring{\lstinline!@coerce!}{@coerce} Statement}%
\label{r:coerce}
In mathematics one often makes use of canonical embeddings of smaller
domains into larger ones, such as $4=4/1$ ($\mathds{Z}\rightarrow\mathds{Q}$)
or $1=1+0i$ ($\mathds{R}\rightarrow\mathds{C}$). In type systems the
according concept are implicit conversions or so-called \emph{coercions}.
In \haggies{} a very primitive implementation of coercions exists:
whenever a type different from the expected is found, \haggies{}
scans the list of defined coercions for an exact match and applies
the conversion in case of a success. The coercions are simple in the
sense that if coercions $c_1:T_1\rightarrow T_2$ and $c_2:T_2\rightarrow T_3$
are defined, \haggies{} will not automatically apply $c_2\circ c_1$
where $T_1$ is found and $T_3$ is expected. Instead, the user has to
define a coercion $T_1\rightarrow T_3$ explicitly.

Coercions can have an optional pattern which specifies an appropriate
conversion function. Consider the examples below:
\begin{lstlisting}
@coerce
   @int -> Int;
   @int -> Float = "%s.0";
   @int/@int -> Float = "%s/%s.0";
   Float -> Polynomial = "(new Polynomial(%s, 0));
\end{lstlisting}
If a pattern is used the wildcard contains the value to be converted.
A special case is the type \lstinline!@int/@int! where two wildcards
are available, one for the numerator and a second one for the denominator.
In the last line of the example we assume the existence of a class
\texttt{Polynomial} with a constructor
\texttt{Polynomial(float coeff, int power)} that constructs the
term $\mathtt{coeff}\cdot x^{\mathtt{power}}$.

\subsection{The \texorpdfstring{\lstinline!@operator!}{@operator} Statement}%
\label{r:operator}
Each multiplicative operator including dot-products and powers
must be defined by an \lstinline!@operator! statement.\footnote{
Powers by integer exponents are implicit if the operator
\lstinline!T*T->T! is defined. Addition operators are implicitly
defined by the \lstinline!@type! statement.}

The \lstinline!@operator! statement can have one or more right-hand
sides, depending on the operator.
\begin{lstlisting}
@operator <+$T_1$+> * <+$T_2$+> -> <+$T_R$+> = <+
$P(ab)$+>, <+$P(a/b)$+>, <+$P(b/a)$+>;
@operator <+$T_1$+> . <+$T_2$+> -> <+$T_R$+> = <+$P(u\cdot v)$+>;
@operator <+$T_1$+> ^ <+$T_2$+> -> <+$T_R$+> = <+$P(u^v)$+>;
\end{lstlisting}
Here, $P(\ldots)$ denotes patterns for the according operations.
In case of the multiplication they default to
\lstinline!"%s*%s"!, \lstinline!"%s/%s"!
and \lstinline!"%2$s/%1$s"! respectively for most languages.
The last pattern is used only if $T_1$ and $T_2$ are different.
Care has to be taken because in $P(b/a)$ the order of the arguments is
not reversed automatically.

The default for exponentiation operator is defined according to the
target language as \lstinline!"%s^%s"!, \lstinline!"%s**%s"! or
\lstinline!"pow(%s,%s)"!.
For languages without a default dot-product the default for the second
line is \lstinline!"__DOT__(%s,%s)"! to indicate the missing
definition in the output.

\subsection{The \texorpdfstring{\lstinline!@nullary!}{@nullary} Statement}%
\label{r:nullary}
The configuration file may contain zero, one or more \lstinline!@nullary!
statements.
This statement declares a list of functions that take zero
arguments. In some languages and \acp{CAS} no distinction is made
between $f$ and $f()$. For optimisation purposes it might, however,
be necessary to detect symbols which require a function call and
to evaluate them only once. For example, some languages implement
the number $\pi$ as a nullary function call rather than a predefined
constant. The declaration
\begin{lstlisting}
@nullary pi;
\end{lstlisting}
would ensure that the function \lstinline!pi! is called only once
and assigned to a symbol during \ac{CSE}. The configuration file may
contain zero, one or more occurrences of the \lstinline!@nullary!
statement.

\subsection{The \texorpdfstring{\lstinline!@polynomial!}{@polynomial} %
Statement}\label{r:polynomial}
The configuration file may contain zero, one or more \lstinline!@polynomial!
(or short: \lstinline!@poly!) statements. The keyword must be followed
by a list of symbols, which are factored out during the Horner scheme phase.
All other symbols are considered part of the coefficient of the terms.
No patterns are allowed in this statement.

The following statements
together define the symbols \texttt{s1}, \texttt{s2} and \texttt{s3} as
symbols in the polynomial part of an expression.
\begin{lstlisting}
@polynomial s1, s2;
@poly s3;
\end{lstlisting}

\subsection{The \texorpdfstring{\lstinline!@brackets!}{@brackets} Statement}%
\label{r:brackets}
There are two forms of bracketing in \haggies{}. The first
form has the syntax
\begin{lstlisting}
@brackets <+$number$+>;
\end{lstlisting}
where $number$ must be a positive integer literal. This statement ensures
that no more than $number$ terms are grouped together and calculated in a
subprogram. In the template file the loop \lstset{language=[c]{haggiest}}
\lstinline![% @for brackets %]!
can be used to enumerate all brackets generated by this statement.

In the second form of the \lstset{language=haggies} the \lstinline!@brackets!
keyword is followed by a list of patterns. Those patterns are factored out
and the remaining factors are collected inside brackets and computed in a
subprogram. If a pattern appears inside parenthesis it matches any factor that
contains a symbol matching this pattern; in particular,
factors are matched that contain this symbol inside function arguments
or in denominators.
Without parenthesis patterns are not recognised inside function arguments.

An example of a valid brackets statement is the following.
\begin{lstlisting}
@brackets a, b...;
@brackets f, (x), y, (...b);
\end{lstlisting}
With this declaration, \haggies{} factors out the symbols `\texttt{a}'
and `\texttt{y}' and all symbols that start with the letter `\texttt{b}'.
It also
factors out all occurrences of the function \texttt{f}\footnote{assuming
\texttt{f} was declared as a function} and all factors that contain
either directly or as part of a subexpression
the symbol \texttt{x} or symbols ending with \texttt{b}.

Brackets can be nested arbitrarily. The level is indicated by a
number in square brackets following directly after the initial keyword.
The following example defines three levels of brackets.
\begin{lstlisting}
@brackets [1] s12, s23;
@brackets [10] 10000;
@brackets [5] (x1), (x2), Log;
@brackets [1] m...;
\end{lstlisting}
At the first level the variables \texttt{s12}, \texttt{s23} and all
symbols starting with an \texttt{m} are selected. The second level
of brackets factors out all occurrences of \texttt{x1} and \texttt{x2}
and all logarithms. The innermost level ensures that no more than 10,000
terms are computed in a subprogram.

It should be noted that neither the order of the \lstinline!@brackets!
statements nor the value of the bracket level is important but only
the relative ordering of the levels ($1<5<10$). The two forms of bracketing
(by number or by pattern) cannot be mixed at the same level of nesting.
Nested brackets can be accessed through nested
\lstset{language=[c]{haggiest}}\lstinline![% @for brackets %]!
loops in the template file.

As a word of warning we like to add that excessive bracketing can decrease
the degree of optimisation as no optimisation can be applied
across the different brackets.

\section{Templates}\label{ssec:reference:out}
\lstset{language=[c]{haggiest}}
\subsection{Markup Structure}
The structure of the template files is closely related to other
markup languages such as XML. However, since `$\langle$' and `$\rangle$'
are already heavily used in most target languages their use for indication
of markup tags would lead to a proliferation of their escaped representation
\footnote{e.g. \texttt{\&lt;} and \texttt{\&gt;} in XML}. Since this is
true for virtually any character we use the two letter combinations
\lstinline![% ... %]!.

The first word inside the brackets must be either a keyword
preceded by an `\texttt{@}' sign or the name of a predefined function.
The remainder of the tag consists of either single words indicating flags
or of options of the form \lstinline!key="value"! or \lstinline!key=value!
(with or without quotes). Quoted strings can contain the usual
backslash-escape sequences. In addition, the sequence \lstinline!"\N"!
creates a line separator according to the operating system standard.\footnote{
In fact the \texttt{Java} property \texttt{line{.}separator} is used.}

Inside a tag, a comment is indicated by a single quote (\texttt{'}) and terminates with
the end of the tag.

\subsection{Control Structures}
Control structures always start
with a keyword; the corresponding end-tag has the form \lstinline![% @end @keyword %]!.
Arguments in the end-tag are always ignored and can be used as commentary.

\subsubsection{Conditional Branching}\label{r:if}\label{r:select}
There are two control structures that can be used for the selection of conditional branches
in a template file. The if--elif--else structure and the select--case--else block.

The select--case--else block requires at least one case-branch; the else-branch is
optional. The first case-branch is inside the select-tag.
\begin{lstlisting}[language={[c]haggiest}]
[% @select <+$name$%
   +> <+$opt_{1}[$+>=<+$value_{1}]$+> <+$opt_{2}[$+>=<+$value_{2}]$+> <+%
   $\ldots$+>

   @case <+$value_{11}$+> <+$value_{12}$+> <+$\ldots$+>
      %]branch 1[%
   @case <+$value_{21}$+> <+$value_{22}$+> <+$\ldots$+>
      %]branch 2[%
      <+$\vdots$+>
   @else
      %]branch n[%
   @end @select %]
\end{lstlisting}
Its semantics is as follows: The expression in the first tag is evaluated
with its semantic as a function. The result is compared to the values in
the first case-tag. If one of the values matches the result
the according branch is evaluated and evaluation continues after
the select--case--else block.
Otherwise the
following case-branches are tested until either one of the tests succeeds
or the optional else-branch is reached. Unlike the switch-statement in
\texttt{C}-like languages, the select--case--else block ensures that at
most one branch is evaluated.
If no else-branch is present and all tests fail the block evaluates to
an empty string.
It should be noted that the values in the case-branches are not evaluated
further, i.e.
are not interpreted as variable or function names but taken as literal values.

The if--elif--else structure contains at least one branch
(here: \lstinline!branch 1!).
All elif-branches and the else-branch are optional.
\begin{lstlisting}[language={[c]haggiest}]
[% @if <+$name_1$%
   +> <+$opt_{1,1}[$+>=<+$value_{1,1}]$+> <+$opt_{1,2}[$+>=<+$value_{1,2}]$+> <+$\ldots$+>
      %]branch 1[%
   @elif <+$name_2$%
   +> <+$opt_{2,1}[$+>=<+$value_{2,1}]$+> <+$opt_{2,2}[$+>=<+$value_{2,2}]$+> <+$\ldots$+>
      %]branch 2[%
      <+$\vdots$+>
   @else
      %]branch n[%
   @end @if %]
\end{lstlisting}
The expression in the if-branch is evaluated and the result converted to a truth-value.
If the result of the evaluation is \textit{true} the according branch is evaluated;
otherwise the elif-branches are processed in this way until the first elif-branch
returns \textit{true}. If no test in the elif-branches succeeded the optional else-branch
is evaluated. If no else-branch is present and all tests fail this structure evaluates to
the empty string.

\subsubsection{Repetition}\label{r:for}
Repeated evaluation of a block can be achieved with the for-loop.
\begin{lstlisting}[language={[c]haggiest}]
[% @for <+$name$%
   +> <+$opt_{1}[$+>=<+$value_{1}]$+> <+$opt_{2}[$+>=<+$value_{2}]$+> <+$\ldots$+>
      %]loop body[%
   @end @for %]
\end{lstlisting}
The $name$ must be the name of a predefined iterator
(See Section~\ref{ssec:iterators}).
For each element of the iterator
a set of variables (depending on the iterator) is set to the according values and made
visible within the loop body. The loop body is evaluated and appended to the output
for each iteration.

\subsubsection{Scopes}\label{r:with}
The with-statement allows to retrieve a set of variables from a source and to make them
visible within its scope.
\begin{lstlisting}[language={[c]haggiest}]
[% @with <+$name$%
   +> <+$opt_{1}[$+>=<+$value_{1}]$+> <+$opt_{2}[$+>=<+$value_{2}]$+> <+$\ldots$+>
	   %]body[%
   @end @with %]
\end{lstlisting}
The $name$ must be the name of a predefined environment
(See Section~\ref{ssec:environments}).
A set of variables is set to the according values and made visible within the scope of the
with statement.

\subsection{Predefined Iterators}\label{ssec:iterators}
\subsubsection{repeat}\label{i:repeat}
Repeats the loop body a given number of times.
The options for this iterator are:
\begin{description}
\item[from] (default=1) start value
\item[to] end value
\item[by] (default=1) increment
\item[shift] (default=0) value added to the counter when assigned to
`\$\_'
\item[prefix] (optional) a common prefix to all variable names
declared by this iterator.
\end{description}

Inside the loop the following variables are defined:
\begin{description}
\item[is\_first] Boolean value which indicates if this is the
first iteration of the loop
\item[is\_last] Boolean value which indicates if this is the
last iteration of the loop
\item[\$\_] the current value of the counter
\end{description}

\subsubsection{brackets}\label{i:brackets}
Iterates over all brackets; can be nested to
    access nested levels of brackets.
The options for this iterator are:
\begin{description}
\item[prefix] (optional) a common prefix to all variables defined
    by this iterator.
\item[only] (optional) expects a type name as value. If present,
    only brackets of the given return type are enumerated.
\item[bracket] \texttt{printf} format which is used to format
    the bracket symbol.
    For each level of nesting an extra format field can be used. The last
	 format field corresponds to the most deeply nested bracket.
	 Hence for the third subbracket of the fifth bracket the
	 string \lstinline!"%2$d_%1$d"! generates `5\_3'.
\item[result] determines how the variable \lstinline!&result!
    is formatted.
\end{description}

Inside the loop the following variables are defined:
\begin{description}
\item[is\_first] Boolean value which indicates if this is the
first iteration of the loop
\item[is\_last] Boolean value which indicates if this is the
last iteration of the loop
\item[\$\_] the current bracket symbol formatted by the `bracket'
option.
\item[index] the index of the most deeply nested bracket
\item[type.name] the name of the data type of this bracket
\item[type.repr] the representation of the data type of this bracket
\end{description}

\subsubsection{instructions}\label{i:instructions}
Enumerates the instructions of the
current subprogram in the correct order.

The options for this iterator are:
\begin{description}
\item[prefix] (optional) a common prefix to all variables defined
    by this iterator.
\end{description}

Inside the loop the following variables are defined:
\begin{description}
\item[is\_first] Boolean value which indicates if this is the
first iteration of the loop
\item[is\_last] Boolean value which indicates if this is the
last iteration of the loop
\item[is\_zero] Boolean value which indicates if the right-hand
side is identically zero.
\item[index] the index of this instruction
\item[\$\_] the current left-hand side symbol
\item[type.name] the name of the data type of the left-hand side
of this assignment
\end{description}

\subsubsection{symbols}\label{i:symbols}
Enumerates all temporary symbols of this
subroutine.

The options for this iterator are:
\begin{description}
\item[prefix] (optional) a common prefix to all variables defined
    by this iterator.
\item[only] (optional) expects a type name as value. If present,
    only symbols of the given return type are enumerated.
\end{description}

Inside the loop the following variables are defined:
\begin{description}
\item[is\_first] Boolean value which indicates if this is the
first iteration of the loop
\item[is\_last] Boolean value which indicates if this is the
last iteration of the loop
\item[\$\_] the current symbol
\item[type.name] the name of the data type of this symbol
\item[type.repr] the representation of the data type of this symbol
\end{description}

\subsection{Predefined Environments}\label{ssec:environments}
It should be noted, that wherever it is possible to define new variable
names the user must not use `path' and `IP' as names
as these are used internally.

\subsubsection{env} If the option `file' is specified it must point to
    an existing file, from which a set of properties is read and
	 assigned to variables. Otherwise all options are interpreted
	 as variable assignments.
\subsubsection{args} Provides all command line definitions (\texttt{-D} options)
    as variables. The option `prefix' can be used to put a common prefix
	 in front of all variable names.
\subsubsection{os} Provides all environment variables of the shell or
    the operating system as variables.
	 The option `prefix' can be used to put a common prefix
	 in front of all variable names.
\subsubsection{vm} Provides all system properties of the \texttt{Java}
    virtual machine as variables.
	 The option `prefix' can be used to put a common prefix
	 in front of all variable names.
\subsubsection{eval} Similar to the `env' environment with the difference
    that the right-hand sides are evaluated as described in
	 Section~\ref{sssec:calculator}.

\subsection{Predefined Functions}

\subsubsection{time\_stamp} Returns the current time and date.
\begin{description}
\item[format] (optional) specifies the output format according to the
              rules used in \texttt{java.text.SimpleDateFormat}.
\item[locale] (optional) specifies a language and/or a country
              for the formatting of the date,
              e.g. `en\_GB', `en\_US', `fr', `de\_CH' \dots
\end{description}
\subsubsection{user\_name} Returns the name of the current user.
This name is determined by the \texttt{Java} property
\texttt{user.name}.
\subsubsection{eval} Evaluates an expression
according to the expression syntax described in
Section~\ref{sssec:calculator}. The result can be transformed
with the same rules as described in Section~\ref{sssec:evaluation}.
\begin{description}
\item[expression] the expression to be evaluated
\end{description}

\subsubsection{LINE} Returns the current line-number in the output file. This can be
used to generate error messages in languages that do not automatically support
exception handling.

\subsection{Other Commands}
The following commands can be used as functions, with the only difference
that they cannot appear in the tags of an if- or select-statement.
\subsubsection{include} Includes a file specified by its file name.
The file is then interpreted as if it was part of the template file.
In particular, all tags in the included file are interpreted
and all variables defined at the point of the include statement
are visible inside the included file.
\begin{description}
\item[file] name of the file to be included
\end{description}

\subsubsection{tab} The tag \lstinline![% tab %]! generates a horizontal tabulator
character\footnote{i.e. '\textbackslash t' = ASCII \texttt{0x09}}.
\begin{description}
\item[rep] (optional) specifies how many tab-characters should be printed.
\end{description}

\subsubsection{nl} The tag \lstinline![% nl %]! generates a new line
character. Which character(s) are printed is determined by the \texttt{Java}
property \texttt{line.separator}.
\begin{description}
\item[rep] (optional) specifies how many new-line sequences should be printed.
\end{description}

\subsubsection{expression} This function is defined only inside the
`instruction' iterator. It generates a suitable representation
for the right-hand side of the current assignment.
\begin{description}
\item[match] Together with the `format' option this is used to
transform the names of the temporary variables (`\$1', `\$2', \dots)
The pattern provided in the option `match' is applied to each symbol
in the expression. If the match fails, the symbol is returned unchanged.
If it succeeds the option `format' is applied to the groups of the match.
\item[format] a \texttt{printf} format. See option `match' for details.
\item[bracket] a \texttt{printf} format used to format bracket names.
For more details see the documentation on the `brackets' iterator.
\end{description}

\subsection{Variable Evaluation}\label{sssec:evaluation}
If no options are specified, variables evaluate to its current value as it is.
However, there are a couple of transformations that can be applied to any
variable and some of the functions by providing a set of options.
\begin{description}
\item[match] (optional) a \texttt{Java} regular expression that
       must match the value of the variable.
\item[format] (optional) a \texttt{printf} format string. If
       the above regular expression is present and contains
		 groups\footnote{Groups are regular expressions in parenthesis.}
       the wildcards are filled with the contents of the groups. Otherwise
		 the value itself is used.
\item[convert] (optional) if present, a conversion is applied to the
       value before it is used in the format string. Possible values are
		 \begin{description}
		 \item[upper] convert to upper case letters
		 \item[lower] convert to lower case letters
		 \item[bool]  interprets value as logical value. If the options
		 `true' and `false' are specified their values are used instead;
       example:
\begin{lstlisting}
[% iszero convert=boolean
          true=ZERO false=NON_ZERO%]
\end{lstlisting}
		 \item[number] interprets the value as a number. If the option
		 `radix' is specified the value is assumed to be in the given
		 radix.
		 \end{description}
\item[locale] (optional)
   a description of a country and/or language separated by
	underscores. This string is specified by the ISO two-letter
	combinations for countries and languages (e.g.
	`en\_US', `de\_DE', `fr\_CA' or language only `en', `de', `fr').
\item[default] (optional) value to be used if a variable is not defined
\end{description}
Since internally all variables are stored as strings, conversion to
numbers is required if one wants to apply number format descriptors
in the format. An example would be
\lstinline![% helicity format="%7d" convert=number %]!.
\subsection{The Built-in Calculator}\label{sssec:calculator}
The predefined environment and function `eval' uses the following set of
operators and functions. Function names must be quoted in single quotes
to avoid their interpretation as variable names unless the function name
should be retrieved from a variable.

Table~\ref{tbl:operator-prec}
collects all implemented operators in their precedence from
strong to weak operator binding. Table~\ref{tbl:eval-functions}
contains a list of all implemented functions. The term regex refers
to a string constant that uses \texttt{Java} regular expression
format.
\begin{longtable}[c]{lp{9cm}}
\caption{Operator precedence of the `\lstinline!eval!' function}%
\label{tbl:operator-prec}%
\endfirsthead
\multicolumn{2}{c}{Table~\ref{tbl:operator-prec}:
Operator precedence of the `\lstinline!eval!' function
\it (continued)}\\[1ex]
\endhead
\multicolumn{2}{c}{\it Table is continued on the next page.}
\endfoot
\endlastfoot
\hline
\lstinline!s[m:n]!  & extract substring (zero-based)\\
\lstinline!x(...)!  & function call (see extra table)\\
\lstinline|x!|      & factorial\\
\lstinline|x!!|     & double factorial\\
\hline
\lstinline!=x!      & test: is not null\\
\lstinline!#x!      & string length\\
\lstinline!*x!      & dereference: interpret as variable name\\
\lstinline!~x!      & convert integer to float\\
\lstinline|!x|      & logical not\\
\lstinline!|x!      & magnitude\\
\lstinline!<>x!     & trim: remove leading and trailing blanks\\
\lstinline!<<x!     & left trim: remove leading blanks\\
\lstinline!>>x!     & right trim: remove trailing blanks\\
\hline
\lstinline!x^y! or \lstinline!x**y! & power\\
\lstinline!s@n!     & remove leftmost $n-1$ characters from string\\
\lstinline!s|n!     & retain only leftmost $n$ characters of string\\
\lstinline!s#t!     & concatenate strings\\
\lstinline!s^^n!    & repeat string $n$ times\\
\hline
\lstinline!+x!      & unary plus\\
\lstinline!-x!      & unary minus\\
\hline
\lstinline!x*y!     & arithmetic multiplication\\
\lstinline!x/y!     & arithmetic (true) division\\
\lstinline!x//y!    & integer division\\
\lstinline!x%y!     & remainder of integer division\\
\hline
\lstinline!x+y!     & arithmetic sum\\
\lstinline!x-y!     & arithmetic difference\\
\lstinline!s<<n!    & align left: append up to $n$ blanks\\
\lstinline!s>>n!    & align right: prepend up to $n$ blanks\\
\lstinline!s<>n!    & centre using a $n$ as width\\
\lstinline!x<:y!    & minimum: lesser of two numbers\\
\lstinline!x<:y!    & maximum: greater of two numbers\\
\hline
\lstinline!s=t!     & equality: compare as strings\\
\lstinline!x==y!    & equality: compare as numbers\\
\lstinline!x!=y!    & inequality: compare as numbers\\
\lstinline!x>=y! etc. & comparison: compare as numbers\\
\hline
\lstinline!x&&y!    & logical and\\
\hline
\lstinline!x||y!    & logical or\\
\lstinline!x><y!    & logical exclusive or\\
\hline
\lstinline!x?s:t!   & ternary if\\
\hline
\end{longtable}

\begin{longtable}[c]{lp{9cm}}
\caption{Functions available in the `\lstinline!eval!' function}%
\label{tbl:eval-functions}%
\endfirsthead
\multicolumn{2}{c}{Table~\ref{tbl:eval-functions}:
Functions available in the `\lstinline!eval!' function
\it (continued)}\\[1ex]
\endhead
\multicolumn{2}{c}{\it Table is continued on the next page.}
\endfoot
\endlastfoot
\lstinline!'cos'(x)!  & trigonometric function\\
\lstinline!'sin'(x)!  & trigonometric function\\
\lstinline!'tan'(x)!  & trigonometric function\\
\lstinline!'acos'(x)!  & inverse trigonometric function\\
\lstinline!'asin'(x)!  & inverse trigonometric function\\
\lstinline!'atan'(x)!  & inverse trigonometric function\\
\lstinline!'atan2'(x,y)! & angle (in radians) between point $(x,y)$
                            and x-axis \\
\lstinline!'exp'(x)!  & $e^x$ \\
\lstinline!'pow'(a,x)!  & $a^x$ \\
\lstinline!'log'(x)!  & $\log x$ \\
\lstinline!'log'(x, a)!  & $\log_ax$ \\
\lstinline!'sqrt'(x)!  & $\sqrt{x}$ \\
\lstinline!'ceil'(x)!  & round towards $+\infty$ \\
\lstinline!'floor'(x)!  & round towards $-\infty$ \\
\lstinline!'round'(x)!  & round towards nearest neighbour
                   $=\mathrm{floor}(x+1/2)$ \\
\lstinline!'sign'(x)!  & sign of $x$\\
\lstinline!'random'(x,y)! & uniformly distributed pseudo-random number
                   between $x$ and $y$\\
\lstinline!'choose'(a,b,c,...)! & pseudo-random selection of one argument\\
\lstinline!'subst'(p,r,s)! & substitute all occurrences of regex $p$ in $s$
                             by $r$ \\
\lstinline!'matches'(p,s)! & test if $s$ matches regular expression $p$\\
\lstinline!'startswith'(s,t)! & test if $s$ starts with prefix $t$\\
\lstinline!'endswith'(s,t)! & test if $s$ ends with suffix $t$\\
\lstinline!'contains'(s,t)! & test if $s$ contains $t$\\
\lstinline!'pos'(t,s)! & first position of $t$ in $s$ (0-based) or $-1$\\
\lstinline!'rpos'(t,s)! & last position of $t$ in $s$ (0-based) or $-1$\\
\lstinline!'lc'(s)! & convert to lower case\\
\lstinline!'uc'(s)! & convert to upper case\\
\lstinline!'words'(s)! & count words separated by white-space\\
\lstinline!'words'(s,i)! & return $i$-th word of $s$ (first word: $i=1$)\\
\lstinline!'split'(s,d)! & count tokens separated by regex $d$\\
\lstinline!'split'(s,d,i)! & return $i$-th token of $s$\\
\lstinline!'format'(f,x,y,...)! & apply \texttt{printf} format sequences
                                  in $f$ to $x$, $y$, \dots\\
\lstinline!'escape'(s,f)! & escape special characters in $s$\\
 & $`e'\in f$: use XML/HTML entities\\
 & $`d'\in f$: escape double quotes\\
 & $`s'\in f$: escape single quotes (by doubling ')\\
 & $`b'\in f$: use backslash sequences\\
 & $`x'\in f$: allow `\lstinline!\x..!' hexadecimal sequences\\
 & $`o'\in f$: allow octal backslash sequences\\
 & $`u'\in f$: use URL encoding
\end{longtable}

\subsection{Emulating Procedure Calls}
The combination of environments with include files allows one
the efficient emulation of procedures, i.e. of repeatedly used,
parametrised text blocks. Consider the following example: first
we write an include file that defines a record in a Pascal-like language.
\begin{lstlisting}[title="foo.prc"]
(* [%f%] is a record with [%n%] entries *)
type [%f%] = record[% @for repeat to=n %]
   field[%$_%] : integer;[%
   @end @for%]
end;
\end{lstlisting}
The second fragment shows its usage.
\begin{lstlisting}
[% @with env f="bar" n="3" %][%
   include file="foo.prc"%][%
   @end with%]
[% @with env f="baz" n="10" %][%
   include file="foo.prc"%][%
   @end with%]
\end{lstlisting}

\subsection{Escaping \texttt{[\%}}
In the unlikely case that one needs to write \texttt{[\%} verbatim in
the output we recommend the use of the following structure.
\begin{lstlisting}
[% @with env ESC="[%" %][% ESC %][% @end @with %]
\end{lstlisting}
The same technique can be used to generate other symbols that are
not or not easy to generate in the input file, e.g.
\begin{lstlisting}
[% @with env uuml="\u00fc" ctrl_L="\x0C" %]
\end{lstlisting}
would define the Unicode character \texttt{uuml}=`\"u'
and the ASCII code Ctrl+L.

\section{Command Line Options}\label{ssec:reference:cli}
\subsection{\texttt{--help} or \texttt{-h}}
    Prints the list of available options and exits.
\subsection{\texttt{--version} or \texttt{-v}}
    Prints the version of the program and exits.
\subsection{\texttt{--verbose[=<value>]} or \texttt{-V [<value>]}}
    Filters which messages are written to the screen.
    If the value is ommitted all status messages with timings are
    printed. The option \texttt{-V2} restricts the output to the
    operation counts.
\subsection{\texttt{--stdin} or \texttt{-f}}
    Adds the standard input to the list of input files.
    If no input files are given at all, this option is implied.
\subsection{\texttt{--template=<value>} or \texttt{-t <value>}}
    Selects the template file. This option must not be ommitted.
\subsection{\texttt{--config=<value>} or \texttt{-c <value>}}
    Selects the configuration file. This option must not be ommitted.
\subsection{\texttt{--output=<value>} or \texttt{-o <value>}}
    Sets the output file. If this option is not specified the output
    file is written to the standard output.
\subsection{\texttt{--cse-on-numbers} or \texttt{-n}}
    Treat numbers as symbols in \ac{CSE}. This option is recommended if
    the data type for numeric constants cannot be represented by literals
    but requires a function or constructor call. If the \texttt{-n} option
    is specified the program introduces variables for all numeric constants
    to save constructor calls.
\subsection{\texttt{--horner=<value>} or \texttt{-H <value>}}
    Selects a strategy for the horner scheme. The default is the
    original greedy strategy, which is implemented in the class
    \texttt{SingleCount}.
    The value must correspond to a class name
    in the package \texttt{haggies.analyser.strategies}.
    Optionally, one can add arguments for the constructor separated by
    colons. Please, consult the API documentation for more details.
    If the class name is \texttt{null} no Horner scheme is carried out.
\subsection{\texttt{--allocator=<value>} or \texttt{-A <value>}}
    Selects an algorithm for the variable allocation. Possible values
    are
    \begin{enumerate}
    \item['C'] selects the graph colouring algorithm~\cite{briggs98register}
        for
        variable allocation. This algorithm typically fails for large
        expressions.
     \item['E'] selects the Extended Linear Search strategy. This option
        is the default.
     \item['S'] an alternative implementation of the Extended Linear
        Search algorithm with reduced memory usage but slightly slower.
    \end{enumerate}
\subsection{\texttt{--no-expand} or \texttt{-E}}
    If this option is set the expansion of terms in parentheses is
    suppressed. This option should be used if the input expression
    is provided in a (partially) factorised form which should be kept
    in the output.
\subsection{\texttt{--expand} or \texttt{-e}}
    Expansion of parentheses is enforced. This is the default.


\section{Acronyms}
\begin{acronym}[EBNF]
\acro{AST}{Abstract Syntax Tree}
\acro{CAS}{Computer Algebra System}
\acro{CSE}{Common Subexpression Elimination}
\acro{DAG}{Directed Acyclic Graph}
\acro{EBNF}{Extended Backus-Naur Form}
\acro{JAR}{\texttt{Java} Archive}
\acro{JDK}{\texttt{Java} Development Kit}
\acro{LHC}{Large Hadron Collider}
\acro{QCD}{Quantum Chromodynamics}
\end{acronym}

\bibliography{haggies-cpc}
\end{document}

%% file: lstform.tex
\lstdefinelanguage{form}{%
   sensitive=false,
   morecomment=[f][commentstyle][0]{*},
   morecomment=[f][keywordstyle][0]{.},
   morestring=[b]",
   morecomment=[n][stringstyle]{`}{'},
   moredelim=*[directive]\#,
   moredirectives={append,break,call,case,close,commentchar,create,default},
   moredirectives={define,do,else,elseif,enddo,endif,endprocedure,endswitch},
   moredirectives={exchange,external,fromexternal,if,ifdef,ifndef,include},
   moredirectives={message,pipe,preout,procedure,procedureextension,prompt},
   moredirectives={redefine,remove,rmexternal,setexternal,setexternalattr},
   moredirectives={show,switch,system,terminate,toexternal,undefine,write,},
   morekeywords={global, local},
   morekeywords={symmetric, antisymmetric, cyclesymmetric},
   morekeywords={rcyclesymmetric, rcyclic, reversecyclic},
   morekeywords={symbol, symbols, cfunction, cfunctions},
   morekeywords={function, functions, vector, vectors},
   morekeywords={tensor, tensors, ctensor, ctensors},
   morekeywords={set, sets, index, indices, table, ctable},
   morekeywords={dimension, dimensions, unittrace},
   morekeywords={if, else, elseif, endif, while},
   morekeywords={repeat, endrepeat, label, goto},
   morekeywords={argument, endargument, exit},
   morekeywords={inexpression, inside, term},
   morekeywords={endinexpression, endinside, endterm},
   morekeywords={abracket, antibracket, bracket},
   morekeywords={abrackets, also, antibrackets, antisymmetrize},
   morekeywords={argexplode, argimplode, apply, auto, autodeclare},
   morekeywords={brackets, chainin, chainout, chisholm, cleartable},
   morekeywords={collect, commuting, compress, contract},
   morekeywords={cyclesymmetrize, deallocatetable, delete},
   morekeywords={dimension, discard, disorder, drop, factarg, fill},
   morekeywords={fillexpression, fixindex, format, funpowers, hide},
   morekeywords={id, identify, idnew, idold, ifmatch, inparallel},
   morekeywords={insidefirst, keep, load, makeinteger, many, metric},
   morekeywords={moduleoption, modulus, multi, multiply, ndrop},
   morekeywords={nfunctions, nhide, normalize, notinparallel},
   morekeywords={nprint, nskip, ntable, ntensors, nunhide, nwrite},
   morekeywords={off, on, once, only, polyfun, pophide, print},
   morekeywords={printtable, propercount, pushhide, ratio},
   morekeywords={rcyclesymmetrize, redefine, renumber},
   morekeywords={replaceinarg, replaceloop, save, select},
   morekeywords={setexitflag, skip, slavepatchsize, sort, splitarg},
   morekeywords={splitfirstarg, splitlastarg, sum, symmetrize},
   morekeywords={tablebase, testuse, threadbucketsize, totensor},
   morekeywords={tovector, trace4, tracen, tryreplace, unhide},
   morekeywords={unittrace, vectors, write},
   morekeywords={abs_, bernoulli_, binom_, conjg_, count_},
   morekeywords={d_, dd_, delta_, deltap_, denom_, distrib_},
   morekeywords={dum_, dummy_, dummyten_, e_, exp_, fac_},
   morekeywords={factorin_, firstbracket_, g5_, g6_, g7_},
   morekeywords={g_, gcd_, gi_, integer_, invfac_, match_},
   morekeywords={max_, maxpowerof_, min_, minpowerof_},
   morekeywords={mod_, nargs_, nterms_, pattern_, poly_},
   morekeywords={polyadd_, polydiv_, polygcd_, polyintfac_},
   morekeywords={polymul_, polynorm_, polyrem_, polysub_},
   morekeywords={replace_, reverse_, root_, setfun_, sig_},
   morekeywords={sign_, sum_, sump_, table_, tbl_, term_},
   morekeywords={termsin_, termsinbracket_, theta_, thetap_},
   morekeywords={5_, 6_, 7_},
   morekeywords={sqrt_, ln_, sin_, cos_, tan_, asin_, acos_},
   morekeywords={atan_, atan2_, sinh_, cosh_, tanh_, asinh_},
   morekeywords={acosh_, atanh_, li2_, lin_}
}[keywords,strings,comments,directives]

%% file: lsthaggies.tex
\lstdefinelanguage{haggies}{%
	sensitive=false,
	morecomment=[l][commentstyle]\#,
	morestring=[b]",
	moredelim=*[directive]@,
	moredirectives={language,type,operation,coerce,define,operator},
	moredirectives={polynomial,nullary,poly,int,brackets},
	morekeywords={@int},
	literate={->}{$\rightarrow$}1
}[directives,strings,comments]

\lstdefinelanguage{haggiest}{%
	sensitive=false,
	morecomment=[s][\rm]{[\%}{\%]},
	morestring=[b]"
}
\lstdefinelanguage[c]{haggiest}[]{c}{%
	sensitive=false,
	morecomment=[s][\color{red}]{[\%}{\%]},
	morestring=[b]"
}
\lstdefinelanguage[fortran]{haggiest}[95]{fortran}{%
	sensitive=false,
	morecomment=[s][\color{red}]{[\%}{\%]},
	morestring=[b]"
}
\lstdefinelanguage[python]{haggiest}[]{python}{%
	sensitive=false,
	morecomment=[s][\color{red}]{[\%}{\%]},
	morestring=[b]"
}
\lstdefinelanguage[java]{haggiest}[]{java}{%
	sensitive=false,
	morecomment=[s][\color{red}]{[\%}{\%]},
	morestring=[b]"
}

%% file: haggies-cpc.bbl
\begin{thebibliography}{10}

\bibitem{Bern:2008ef}
Bern, Z. et~al.,
\newblock (2008),
\newblock [\href{http://arXiv.org/abs/0803.0494}{\texttt{0803.0494}}].

\bibitem{Berger:2008ag}
Berger, C.~F. et~al.,
\newblock (2008),
\newblock [\href{http://arXiv.org/abs/0807.3705}{\texttt{0807.3705}}].

\bibitem{Binoth:2008uq}
Binoth, T., Guillet, J.~P., Heinrich, G., Pilon, E., and Reiter, T.,
\newblock Computer Physics Communications  (2009),
\newblock [\href{http://arXiv.org/abs/0810.0992}{\texttt{0810.0992}}].

\bibitem{Bredenstein:2008ia}
Bredenstein, A., Denner, A., Dittmaier, S., and Pozzorini, S.,
\newblock (2008),
\newblock [\href{http://arXiv.org/abs/0807.1453}{\texttt{0807.1453}}].

\bibitem{Diakonidis:2009fx}
Diakonidis, T., Fleischer, J., Riemann, T., and Tausk, J.~B.,
\newblock (2009),
\newblock [\href{http://arXiv.org/abs/0907.2115}{\texttt{0907.2115}}].

\bibitem{Fujimoto:2008zz}
Fujimoto, J. and Kurihara, Y.,
\newblock Nucl. Phys. Proc. Suppl. {\bf 183} (2008) 143.

\bibitem{Giele:2008bc}
Giele, W.~T. and Zanderighi, G.,
\newblock JHEP {\bf 06} (2008) 038,
\newblock [\href{http://arXiv.org/abs/0805.2152}{\texttt{0805.2152}}].

\bibitem{Hahn:2000kx}
Hahn, T.,
\newblock Comput. Phys. Commun. {\bf 140} (2001) 418,
\newblock
  [\href{http://arXiv.org/abs/hep-ph/0012260}{\texttt{hep-ph/0012260}}].

\bibitem{Hahn:2006qw}
Hahn, T. and Rauch, M.,
\newblock Nucl. Phys. Proc. Suppl. {\bf 157} (2006) 236,
\newblock
  [\href{http://arXiv.org/abs/hep-ph/0601248}{\texttt{hep-ph/0601248}}].

\bibitem{Lazopoulos:2008ex}
Lazopoulos, A.,
\newblock (2008),
\newblock [\href{http://arXiv.org/abs/0812.2998}{\texttt{0812.2998}}].

\bibitem{Reiter:2009kb}
Reiter, T.,
\newblock (2009),
\newblock [\href{http://arXiv.org/abs/0903.0947}{\texttt{0903.0947}}].

\bibitem{vanHameren:2009dr}
van Hameren, A., Papadopoulos, C.~G., and Pittau, R.,
\newblock (2009),
\newblock [\href{http://arXiv.org/abs/0903.4665}{\texttt{0903.4665}}].

\bibitem{vanHameren:2009vq}
van Hameren, A.,
\newblock (2009),
\newblock [\href{http://arXiv.org/abs/0905.1005}{\texttt{0905.1005}}].

\bibitem{Winter:2009kd}
Winter, J.-C. and Giele, W.~T.,
\newblock (2009),
\newblock [\href{http://arXiv.org/abs/0902.0094}{\texttt{0902.0094}}].

\bibitem{Binoth:2009fk}
Binoth, T.,
\newblock (2009),
\newblock [\href{http://arXiv.org/abs/0903.1876}{\texttt{0903.1876}}].

\bibitem{Catani:1996vz}
Catani, S. and Seymour, M.~H.,
\newblock Nucl. Phys. {\bf B485} (1997) 291,
\newblock
  [\href{http://arXiv.org/abs/hep-ph/9605323}{\texttt{hep-ph/9605323}}].

\bibitem{Gleisberg:2007md}
Gleisberg, T. and Krauss, F.,
\newblock Eur. Phys. J. {\bf C53} (2008) 501,
\newblock [\href{http://arXiv.org/abs/0709.2881}{\texttt{0709.2881}}].

\bibitem{Seymour:2008mu}
Seymour, M.~H. and Tevlin, C.,
\newblock (2008),
\newblock [\href{http://arXiv.org/abs/0803.2231}{\texttt{0803.2231}}].

\bibitem{Hasegawa:2008ae}
Hasegawa, K., Moch, S., and Uwer, P.,
\newblock Nucl. Phys. Proc. Suppl. {\bf 183} (2008) 268,
\newblock [\href{http://arXiv.org/abs/0807.3701}{\texttt{0807.3701}}].

\bibitem{Frederix:2008hu}
Frederix, R., Gehrmann, T., and Greiner, N.,
\newblock (2008),
\newblock [\href{http://arXiv.org/abs/0808.2128}{\texttt{0808.2128}}].

\bibitem{Czakon:2009ss}
Czakon, M., Papadopoulos, C.~G., and Worek, M.,
\newblock (2009),
\newblock [\href{http://arXiv.org/abs/0905.0883}{\texttt{0905.0883}}].

\bibitem{aho-sethi-ullman}
Aho, A.~V., Sethi, R., and Ullman, J.~D.,
\newblock {\em Compilers. Principles, Techniques and Tools},
\newblock Addison Wesley, Reading, Massachusetts, 1st edition, 1986.

\bibitem{Vermaseren:2000nd}
Vermaseren, J. A.~M.,
\newblock (2000),
\newblock
  [\href{http://arXiv.org/abs/math-ph/0010025}{\texttt{math-ph/0010025}}].

\bibitem{Vermaseren:2006ag}
Vermaseren, J. A.~M. and Tentyukov, M.,
\newblock Nucl. Phys. Proc. Suppl. {\bf 160} (2006) 38.

\bibitem{Ceberio:03}
Ceberio, M. and Kreinovich, V.,
\newblock ACM {\bf 38} (2004) 8.

\bibitem{Knuth:TAOCP1}
Knuth, D.~E.,
\newblock {\em The Art of Computer Programming: Fundamental Algorithms},
  volume~1 of {\em The Art of Computer Programming},
\newblock Addison-Wesley, Reading, Massachusetts, 3rd edition, 1997.

\bibitem{Chaitin:81}
Chaitin, G.~J. et~al.,
\newblock Computer Languages {\bf 6} (1981) 47.

\bibitem{Poletto:99}
Poletto, M. and Sarkar, V.,
\newblock ACM Transactions on Programming Languages and Systems {\bf 21} (1999)
  895.

\bibitem{java:Formatter}
\url{http://java.sun.com/j2se/1.5.0/docs/api/index.html?java/util/Formatter.ht%
ml}.

\bibitem{Cash:90}
Cash, J.~R. and Karp, A.~H.,
\newblock ACM Trans. Math. Softw. {\bf 16} (1990) 201.

\bibitem{Bogacki1989321}
Bogacki, P. and Shampine, L.~F.,
\newblock Applied Mathematics Letters {\bf 2} (1989) 321.

\bibitem{Dormand198019}
Dormand, J.~R. and Prince, P.~J.,
\newblock Journal of Computational and Applied Mathematics {\bf 6} (1980) 19.

\bibitem{Fehlberg70}
Fehlberg, E.,
\newblock Computing {\bf 6} (1970) 61.

\bibitem{java:Pattern}
\url{http://java.sun.com/j2se/1.5.0/docs/api/index.html?java/util/regex/Patter%
n.html}.

\bibitem{Binoth:2005ff}
Binoth, T., Guillet, J.~P., Heinrich, G., Pilon, E., and Schubert, C.,
\newblock (2005),
\newblock
  [\href{http://arXiv.org/abs/hep-ph/0504267}{\texttt{hep-ph/0504267}}].

\bibitem{Ossola:2006us}
Ossola, G., Papadopoulos, C.~G., and Pittau, R.,
\newblock Nucl. Phys. {\bf B763} (2007) 147,
\newblock
  [\href{http://arXiv.org/abs/hep-ph/0609007}{\texttt{hep-ph/0609007}}].

\bibitem{Ossola:2007ax}
Ossola, G., Papadopoulos, C.~G., and Pittau, R.,
\newblock JHEP {\bf 03} (2008) 042,
\newblock [\href{http://arXiv.org/abs/0711.3596}{\texttt{0711.3596}}].

\bibitem{openMP}
Dagum, L. and Menon, R.,
\newblock IEEE Comput. Sci. Eng. {\bf 5} (1998) 46.

\bibitem{Hentenryck}
Hentenryck, P.~v., McAllester, D., and Kapur, D.,
\newblock SIAM Journal on Numerical Analysis {\bf 34} (1997).

\bibitem{More:1981}
Mor{\'e}, J., Garbow, B., and Hillstrom, K.,
\newblock ACM Trans. Math. Software {\bf 7} (1981) 136.

\bibitem{briggs98register}
Briggs, P.,
\newblock {\em Register Allocation via Graph Coloring},
\newblock PhD thesis, Rice University, Houston, Texas, 1992.

\end{thebibliography}
